\documentclass[prb,aps,twocolumn,groupedaddress,nofootinbib]{revtex4-2}

\usepackage{graphicx}
\usepackage{dcolumn}
\usepackage{bm}
\usepackage{multirow}

\usepackage{centernot}

\usepackage[normalem]{ulem}
\usepackage{cancel}

\usepackage{amsmath}
\usepackage{amsthm}
\usepackage{amstext}
\usepackage{amssymb}
\usepackage{mathrsfs}
\usepackage{amsfonts}
\usepackage{amsbsy} 
\usepackage{tensor}
\usepackage{physics}

\usepackage{wasysym}

\usepackage{latexsym}
\usepackage[american]{babel}
\usepackage{bbm}
\usepackage[colorlinks=true, citecolor=blue!90!black, linkcolor=blue!90!black, linktocpage=true, urlcolor=red!70!black]{hyperref}

\newcommand*{\figref}[2][]{%
  \hyperref[{#2}]{%
    \ref*{#2}%
    \ifx\\#1\\%
    \else
      \,#1%
    \fi
  }%
}

\usepackage[all,matrix,cmtip]{xy}

\usepackage{csquotes}
\MakeOuterQuote{"}

\usepackage{color}
\definecolor{red}{rgb}{1,0,0}
\definecolor{blue}{rgb}{0,0,1}
\definecolor{dblue}{rgb}{0,0,0.4}
\definecolor{green}{rgb}{0,1,0}
\definecolor{black}{rgb}{0,0,0}
\definecolor{white}{rgb}{1,1,1}
\definecolor{niceBlue}{RGB}{20,10,237}

\usepackage[dvipsnames]{xcolor}

\definecolor{brn}{rgb}{.8,.4,.0}
\definecolor{redo}{rgb}{1,.5,.0}
\definecolor{ddgrn}{rgb}{0,0.4,0}
\definecolor{dgrn}{rgb}{0,0.55,0}
\definecolor{dbl}{rgb}{0,0,0.5}

\usepackage[bbgreekl]{mathbbol}
\newcommand{\one}{\mathbf{1}}

\newcommand{\Z}{\mathbb{Z}}

\newcommand{\R}{\mathbb{R}}

\newcommand{\p}[1]{\prime\,}

\renewcommand{\t}[1]{\widetilde{#1}}

\newcommand{\da}{\dagger}
\newcommand{\<}{\langle}
\renewcommand{\>}{\rangle}

\newcommand{\pp}{\partial} 
\newcommand{\spp}{\slashed{\partial}}

\newcommand{\bpm}{\begin{pmatrix}}
\newcommand{\epm}{\end{pmatrix}}
\newcommand{\bmm}{\begin{matrix}}
\newcommand{\emm}{\end{matrix}}

\newcommand{\cC}{ {\cal C} }

\newcommand{\cK}{ {\cal K} }

\newcommand{\cP}{ {\cal P} } 
\newcommand{\cQ}{\mathcal{Q}} 
\newcommand{\cR}{ {\cal R} } 
 
\newcommand{\cT}{ {\cal T} } 
\newcommand{\cU}{\mathcal{U}}

\newcommand{\cZ}{ {\cal Z} } 

\usepackage{euscript}

\newcommand{\al}{\alpha} 
\newcommand{\bt}{\beta} 
\newcommand{\del}{\delta} 
\newcommand{\Del}{\Delta} 
\newcommand{\eps}{\epsilon} 
 
\newcommand{\ga}{\gamma}

\newcommand{\la}{\lambda} 
\newcommand{\La}{\Lambda}

\renewcommand{\th}{\theta} 
 
\newcommand{\si}{\sigma}

\usepackage{tikz}
\usepackage{tikz-cd}
\usetikzlibrary{arrows}
\usetikzlibrary{intersections}
\usetikzlibrary{shapes.geometric}
\usetikzlibrary{decorations.pathmorphing, patterns,shapes}
\usetikzlibrary{decorations.markings}

\newcommand{\sfC}{\mathsf{C}}
\newcommand{\sfR}{\mathsf{R}}
\newcommand{\sfT}{\mathsf{T}}

\usepackage{enumerate}
\usetikzlibrary{calc}

\usepackage{slashed}

\usepackage{booktabs}
\newcommand\Tstrut{\rule{0pt}{3.2ex}}         % = `top' strut
\newcommand{\ie}{\begin{equation}\begin{aligned}[]}
\newcommand{\fe}{\end{aligned}\end{equation}}

\usepackage{comment}
\usepackage{mathtools}

\renewcommand{\t}[1]{\widetilde{#1}}
\renewcommand{\k}{\mathbf{k}}

\newcommand{\q}{\mathbf{q}}
\renewcommand{\r}{\mathbf{r}}
\newcommand{\x}{\mathbf{x}}
\newcommand{\y}{\mathbf{y}}

\newcommand{\bz}{{\rm BZ}}

\newcommand{\mb}[1]{\mathbf{#1}}
\usepackage{braket}
\newcommand{\trs}{{\mathsf T}}
\newcommand{\uone}{{\rm U}(1)}

\newcommand{\rw}{\r_\mathrm{W}}
\newcommand{\rb}{\r_\mathrm{B}}
\newcommand{\rfl}{{\mathsf R}}

\renewcommand{\bar}[1]{\overline{#1}}

\newcommand{\on}{\mathsf{Ons}_n}

\newcommand{\otwo}{\mathsf{Ons}_2}

\begin{document}

\hfill MIT-CTP/5869, YITP-SB-2025-10

\title{
Parity anomaly from LSM:\\ exact valley symmetries on the lattice
}

\author{Salvatore D. Pace$^{\,\mb{K}}$, Minho Luke Kim$^{\,\mb{K}}$, Arkya Chatterjee$^{\,\mb{K}}$, Shu-Heng Shao$^{\,\mb{K},\mb{K}\bm{'}}$}
\affiliation{$^{\mb K}$Department of Physics, Massachusetts Institute of Technology,
 Cambridge, Massachusetts 02139, USA
 \\
$^{\mb{K}\bm{'}}$Yang Institute for Theoretical Physics, Stony Brook University, Stony Brook 11794, USA}

\begin{abstract}

We show that the  honeycomb tight-binding model hosts an exact microscopic avatar
of its low-energy SU(2) valley symmetry and parity anomaly. Specifically, the SU(2) valley symmetry arises from a collection of conserved, integer quantized charge operators that obey the Onsager algebra. Along with lattice reflection and time-reversal symmetries, this Onsager symmetry has a Lieb-Schultz-Mattis (LSM) anomaly that matches the parity anomaly in the IR. Indeed, we show that any local Hamiltonian commuting with these symmetries cannot have a trivial unique gapped ground state. 
We study the phase diagram of the simplest symmetric model and survey various deformations, including Haldane's mass term, which preserves only the Onsager symmetry. 
Our results place the parity anomaly in ${2+1}$D alongside Schwinger's anomaly in ${1+1}$D and Witten's SU(2) anomaly in ${3+1}$D as 't Hooft anomalies that can arise from the Onsager symmetry on the lattice.

\end{abstract}

\maketitle

\makeatletter 
\def\l@subsection#1#2{}
\makeatother 

\tableofcontents 

\section{Introduction}

A hallmark of modern condensed matter physics is the emergence of relativistic, massless fermions from nonrelativistic lattice models of fermions. The honeycomb tight‑binding model at zero chemical potential is an archetypal example with its infrared (IR) description containing two free, massless Dirac fermions in ${2+1}$D~\cite{S1984}. These Dirac fermions enjoy an SU(2) valley symmetry which has a parity anomaly. Since the seminal works~\cite{S1984, Fradkin:1986pq, H1988}, condensed matter realizations of parity anomalies have attracted considerable interest. A prominent example is the parity anomaly of $\uone$ fermion number symmetry realized by a single Dirac cone. It arises on the surface of a ${3+1}$D topological insulator~\cite{FKM0607699, FK0611341, QHZ08023537} and underlies the unconventional quantization of the Hall conductivity in monolayer graphene~\cite{Jackiw:1984ji, Schakel:1991af, Zheng:2002zz, GS0506575}. Experimental fingerprints of this $\uone$ parity anomaly have been observed in both topological insulators~\cite{MOK210504127} and graphene~\cite{NG0509330,ZTS0509355}.

Parity anomalies were first discovered in high-energy physics~\cite{Niemi:1983rq, Redlich:1983kn, Redlich:1983dv, Alvarez-Gaume:1984zst}. They are a class of mixed anomalies between orientation-reversing spacetime symmetries, e.g., time-reversal and spatial reflections, and an internal $G$ symmetry, e.g., ${G = \mathrm{U}(1)}$~\cite{Niemi:1983rq} or $\mathrm{SU}(2)$~\cite{Redlich:1983kn, Redlich:1983dv}.\footnote{The terminology ``parity anomaly'' is confusing because parity typically means the spacetime transformation ${(t,\vec x) \to (t, -\vec x)}$, which is orientation-preserving in odd spacetime dimensions~\cite{SW160204251, Witten:2016cio}. We will still adopt this terminology to follow the historical precedent.} 
When $G$ is a gauge symmetry, parity anomalies are Adler-Bell-Jackiw (ABJ)-like anomalies where spacetime reflections are classical symmetries that fail to survive quantization. In this paper, however, $G$ will always be a global symmetry. Consequently, any parity anomaly involving $G$ is an 't Hooft anomaly, which implies an obstruction to a symmetric gapped phase with a unique ground state and no topological order.\footnote{Unlike gauge anomalies, 't Hooft anomalies do not signal an inconsistency of a theory. 't Hooft anomalies of invertible global symmetries are characterized by an anomaly inflow theory in one higher dimension. However, this does not imply that the theory with anomalous symmetry must reside on the boundary of the inflow theory to be well-defined. See, e.g.,~\cite{Choi:2023xjw, Seifnashri:2025vhf} for recent discussions on related notions of 't Hooft anomalies, including the relation to gauging.} 

In the conventional condensed matter realizations, parity anomalies are emergent 't Hooft anomalies.
That is, they exist only in the IR effective field theory and involve emergent symmetries absent from the ultraviolet (UV) theory.
It is quite common that emergent symmetries in the IR come with emergent 't Hooft anomalies. There is, however, an alternative possibility: there could be an anomaly on the lattice that becomes the parity anomaly in the IR. This occurs, for example, when Lieb-Schultz-Mattis (LSM) anomalies~\cite{Lieb:1961fr} for crystalline and internal symmetries on the lattice become 't Hooft anomalies of internal symmetries in the IR~\cite{CZ151102263, 2018PhRvB..97e4412J, CRH170503892, MT170707686, CS221112543}. 
In this scenario, the LSM anomaly matches the 't Hooft anomaly. Importantly, when matching anomalies, the UV and IR symmetry groups need not be the same. 
For instance, the Heisenberg quantum spin chain has an LSM anomaly for $\Z$ lattice translations and SO(3) spin rotations 
that becomes an 't Hooft anomaly for a ${\Z_2\times\mathrm{SO}(3)}$ global symmetry in the IR. It is then natural to ask: Can parity anomalies show similar fingerprints on the lattice and arise from microscopic symmetries?

In this paper, we explicitly demonstrate that this is indeed possible. 
We focus on the parity anomaly for the SU(2) valley symmetry in the tight-binding model at zero chemical potential. It is commonly believed that this valley symmetry cannot exist on the lattice because the two valleys are connected by high-energy states. Here, however, we show that this SU(2) valley symmetry is not emergent, but emanates \cite{CS221112543} from exactly conserved operators of the lattice Hamiltonian.  
Interestingly, this lattice avatar of the valley symmetry forms a different non-Abelian symmetry related to the Onsager algebra.\footnote{Onsager introduced this algebra to solve the two-dimensional classical Ising model~\cite{PhysRev.65.117}. In this paper, the Onsager algebra is not used to solve models but instead describes conserved operators of models. We refer the reader to Refs.~\onlinecite{VOF181209091, M210314569, CPS240912220, PCS241218606, X250110837, GT250307708} for a survey of models with conserved charges forming the Onsager algebra.} 
The IR parity anomaly arises from an LSM anomaly of exact lattice symmetries including time-reversal, reflections, and this Onsager symmetry.  
This provides an exact condensed matter realization of the parity anomaly, complementary to the conventional viewpoint of emergent anomalies in Ref.~\onlinecite{S1984}.

\section{Exact lattice valley symmetries}

\begin{figure}[t]
    \centering
    \begin{tikzpicture}[scale=.75, >=Triangle, thick]
        \definecolor{linkColor}{HTML}{bdbdbd}
        \definecolor{delColor}{HTML}{1b9e77}
        \definecolor{aColor}{HTML}{d95f02}
        \definecolor{yColor}{HTML}{e6ab02}
        \node (1) at (8/3,-0.7698-1.5396-0.7698) {};
        \node (2) at (16/3,-0.7698-1.5396-0.7698) {};
        \node (3) at (4/3,-0.7698-1.5396) {};
        \node (4) at (4/3+8/3,-0.7698-1.5396) {};
        \node (5) at (4/3+16/3,-0.7698-1.5396) {}; 
        \node (6) at (4/3,-0.7698) {};
        \node (7) at (4/3+8/3,-0.7698) {};
        \node (8) at (4/3+16/3,-0.7698) {};
        \node (9) at (0,0) {};
        \node (10) at (8/3,0) {};
        \node (11) at (16/3,0) {};
        \node (12) at (8,0) {};
        \node (13) at (0,1.5396) {};
        \node (14) at (8/3,1.5396) {};
        \node (15) at (16/3,1.5396) {};
        \node (16) at (8,1.5396) {};
        \node (17) at (4/3,1.5396+0.7698) {};
        \node (18) at (4/3+8/3,1.5396+0.7698) {};
        \node (19) at (4/3+16/3,1.5396+0.7698) {};
        \draw[linkColor, shorten >= -4pt, shorten <= -4pt, line width = 2pt] (1) -- (3);
        \draw[linkColor, shorten >= -4pt, shorten <= -4pt, line width = 2pt] (1) -- (4);
        \draw[linkColor, shorten >= -4pt, shorten <= -4pt, line width = 2pt] (2) -- (4);
        \draw[linkColor, shorten >= -4pt, shorten <= -4pt, line width = 2pt] (2) -- (5);
        \draw[linkColor, shorten >= -4pt, shorten <= -4pt, line width = 2pt] (3) -- (6);
        \draw[linkColor, shorten >= -4pt, shorten <= -4pt, line width = 2pt] (4) -- (7);
        \draw[linkColor, shorten >= -4pt, shorten <= -4pt, line width = 2pt] (5) -- (8);
        \draw[linkColor, shorten >= -4pt, shorten <= -4pt, line width = 2pt] (6) -- (9);
        \draw[linkColor, shorten >= -4pt, shorten <= -4pt, line width = 2pt] (6) -- (10);
        \draw[linkColor, shorten >= -4pt, shorten <= -4pt, line width = 2pt] (7) -- (10);
        \draw[linkColor, shorten >= -4pt, shorten <= -4pt, line width = 2pt] (7) -- (11);
        \draw[linkColor, shorten >= -4pt, shorten <= -4pt, line width = 2pt] (8) -- (11);
        \draw[linkColor, shorten >= -4pt, shorten <= -4pt, line width = 2pt] (8) -- (12);
        \draw[linkColor, shorten >= -4pt, shorten <= -4pt, line width = 2pt] (9) -- (13);
        \draw[linkColor, shorten >= -4pt, shorten <= -4pt, line width = 2pt] (10) -- (14);
        \draw[linkColor, shorten >= -4pt, shorten <= -4pt, line width = 2pt] (11) -- (15);
        \draw[linkColor, shorten >= -4pt, shorten <= -4pt, line width = 2pt] (12) -- (16);
        \draw[linkColor, shorten >= -4pt, shorten <= -4pt, line width = 2pt] (13) -- (17);
        \draw[linkColor, shorten >= -4pt, shorten <= -4pt, line width = 2pt] (14) -- (17);
        \draw[linkColor, shorten >= -4pt, shorten <= -4pt, line width = 2pt] (14) -- (18);
        \draw[linkColor, shorten >= -4pt, shorten <= -4pt, line width = 2pt] (15) -- (18);
        \draw[linkColor, shorten >= -4pt, shorten <= -4pt, line width = 2pt] (15) -- (19);
        \draw[linkColor, shorten >= -4pt, shorten <= -4pt, line width = 2pt] (16) -- (19);
        \draw[fill=white, draw=black] (1) circle (4pt);
        \draw[fill=white, draw=black] (2) circle (4pt);
        \fill[black] (3) circle (4pt);
        \fill[black] (4) circle (4pt);
        \fill[black] (5) circle (4pt);
        \draw[fill=white, draw=black] (6) circle (4pt);
        \draw[fill=white, draw=black] (7) circle (4pt);
        \draw[fill=white, draw=black] (8) circle (4pt);
        \fill[black] (9) circle (4pt);
        \fill[black] (10) circle (4pt);
        \fill[black] (11) circle (4pt);
        \fill[black] (12) circle (4pt);
        \draw[fill=white, draw=black] (13) circle (4pt);
        \draw[fill=white, draw=black] (14) circle (4pt);
        \draw[fill=white, draw=black] (15) circle (4pt);
        \draw[fill=white, draw=black] (16) circle (4pt);
        \fill[black] (17) circle (4pt);
        \fill[black] (18) circle (4pt);
        \fill[black] (19) circle (4pt);
        \draw[delColor, -{>[scale=.8]}, line width=2pt] (7) -- node[left] {\normalsize $\bm{\del}$} (4);
        \draw[aColor, -{>[scale=.8]}, line width=2pt] (6) -- (13);
        \draw[aColor, -{>[scale=.8]}, line width=2pt] (6) -- (14);
        \node[aColor, xshift=14pt, yshift = -2pt] at (13) {\normalsize $\mb{a_2}$};
        \node[aColor, xshift=-14pt, yshift = -2pt] at (14) {\normalsize $\mb{a_1}$};
        \draw[yColor, densely dashed,-{>[scale=.8]}, line width=2pt] (11) -- (19);
        \draw[yColor, densely dashed,-{>[scale=.8]}, line width=2pt] (19) -- (12);
        \draw[yColor, densely dashed,-{>[scale=.8]}, line width=2pt] (12) -- (11);
        \draw[yColor, densely dashed,-{>[scale=.8]}, line width=2pt] (8) -- (15);
        \draw[yColor, densely dashed, -{>[scale=.8]}, line width=2pt] (15) -- (16);
        \draw[yColor, densely dashed, -{>[scale=.8]}, line width=2pt] (16) -- (8);
        \node[yColor, xshift=-6pt, yshift = 28pt] at (8) {\normalsize $i t'$};
    \end{tikzpicture}
    \caption{The honeycomb lattice is a triangular lattice with a two-site basis. The primitive lattice vectors of the triangular lattice formed by the white (W) sites are denoted as $\mb{a}_1$ and $\mb{a}_2$. The black (B) sites form another triangular lattice displaced from the white lattice by the vector $\bm{\del}$. The yellow dashed arrows indicate the next-to-nearest neighbor term in Eq.~\eqref{HaldaneTerm}.
    }
    \label{fig:honeycomb-lattice}
\end{figure}
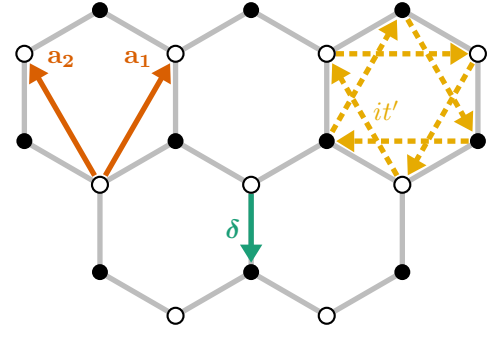

Let us start by considering a honeycomb lattice $\La$ of spinless fermions with a single complex fermion operator $c_\r$ for each site ${\r\in\La}$ satisfying ${\{c_\r, c_{\r'}\} = 0}$ and ${\{c_\r, c^\dag_{\r'}\} = \del_{\r,\r'}}$. 
One of the simplest Hamiltonians for this system is the tight-binding model
\begin{equation}\label{tbHam}
    H 
    =  \sum_{\<\r,\r'\>}
     c^\dag_\r c_{\r'}
    =\sum_{\<\rw,\rb\>} (c^\dag_{\rw} c_{\rb} + c^\dag_{\rb} c_{\rw} ),
\end{equation}
where ${\<\r,\r'\>}$ are nearest neighboring sites. ${\La=\La_\mathrm{W}\oplus \La_\mathrm{B}}$ is bipartite, and we denote by ${\rw\in \La_\mathrm{W}}$ and ${\rb\in\La_\mathrm{B}}$ the lattice sites of its two triangular sublattices (see Fig.~\ref{fig:honeycomb-lattice}). We assume periodic boundary conditions with $L_1$ and $L_2$ sites in the $\mb{a}_1$ and $\mb{a}_2$ directions, respectively.\footnote{Our results and analysis are unchanged when imposing, for instance, anti-periodic boundary conditions. However, for more general twisted boundary conditions, they can change and would depend sensitively on the twisted translation operator.} The total number of sites is $2L_1L_2$. This Hamiltonian is a celebrated toy model known for capturing the band structure of graphene~\cite{GY2019}. It has two Dirac cones, and its continuum limit is two free, massless Dirac fermions in ${2+1}$D.

The Hamiltonian~\eqref{tbHam} commutes with the fermion number operator\footnote{The ${-1/2}$ shift in the definition of $Q$ does not affect its $\uone$ symmetry transformation. It is included such that ${\sfC Q  \sfC^{-1} = -Q}$, where the charge conjugation operator $\sfC$ satisfies~\eqref{latCRT}. Notice that while the eigenvalues of ${c_\r^\dag c_\r - 1/2}$ are $\pm1/2$, the eigenvalues of $Q$ are integer quantized because the honeycomb lattice always has an even number of sites $\r$.}
\begin{equation}\label{FermionNum}
    Q = \sum_{\r\in \La} (c_\r^\dag c_\r - 1/2),
\end{equation}
which has quantized eigenvalues, i.e., ${Q\in \mathbb{Z}}$. The corresponding $\uone$ symmetry operator $e^{i\th Q}$ is on-site and acts on the fermions as ${e^{i\th Q} c_\r e^{-i\th Q} = e^{-i\th} c_\r}$. The model also enjoys various discrete symmetries, including a $\Z_2^{\sfC}$ charge conjugation, $\Z_2^{\sfR}$ reflections, and $\Z_2^{\sfT}$ time-reversal symmetries, which act as:
\ie\label{latCRT}
    \sfC \,c_{\rw} \sfC^{-1} &= c_{\rw}^\dag,
    \qquad
   \sfC \,c_{\rb} \sfC^{-1} = -c_{\rb}^\dag,\\
   \sfR \, c_\r \sfR^{-1} &= c_{\sfR\cdot\r},\\
   \sfT \, c_\r \sfT^{-1} &= c_\r,
    \hspace{35pt} 
    \sfT \, i \, \sfT^{-1} = -i.
\fe
There are six reflection axes of $\La$, which are related by lattice translations and rotations.\footnote{We provide an expanded discussion of the symmetries of the honeycomb lattice in Appendix~\ref{app:crystalSymsHoneycomb}.} We will choose $\sfR$ to be the reflection acting as ${\sfR\cdot  \mb{a}_1= \mb{a}_2}$, ${\sfR\cdot \mb{a}_2 = \mb{a}_1}$, and ${\sfR\cdot \bm{\del} =\bm{\del}}$. The symmetry operators $e^{i\th Q}$, $\sfC$, $\sfR$, and $\sfT$ satisfy $\sfC e^{i\th Q} \sfC^{-1} = \sfR e^{-i\th Q} \sfR^{-1}= \sfT e^{i\th Q} \sfT^{-1} = e^{-i\th Q}$ and furnish a representation of the group ${\mathrm{U}(1)\rtimes(\Z_2^{\sfC}\times \Z_2^{\sfT})\times\Z_2^\sfR}$.
There is another time-reversal symmetry ${\sfT' = \sfC \sfT e^{i\pi Q/2}}$, which squares to the fermion parity $e^{i \pi Q}$ and generates a $\mathbb{Z}_4^\sfT$ symmetry on the lattice.\footnote{There is no parity anomaly between the $\uone$ fermion number $e^{i\th Q}$ and spacetime reflection symmetries. One way to see this is to note that $e^{i\th Q}$, crystalline symmetries, and lattice time-reversal symmetry $\sfT$ are compatible with a trivial unique, gapped ground state. For example, they commute with the Hamiltonian ${\sum_{\r\in\La} c^\dag_\r c_\r}$. The other time-reversal symmetry $\mathbb{Z}_4^\sfT$ generated by $\sfT'$ is also anomaly-free since, for instance, it commutes with the Haldane mass term~\eqref{HaldaneTerm} which gaps out the Dirac cones when added to~\eqref{tbHam}.
}

When studying fermionic models, it is often illuminating to decompose the complex fermion operator into real, Majorana fermion operators. Consider the decomposition
\begin{equation}\label{comptoreal}
    c_{\rw} = \frac12 (a_{\rw} + i b_{\rw}),
    \qquad c_{\rb} = \frac12 (b_{\rb} - i a_{\rb}),
\end{equation}
where the Majorana operators ${a_\r = a_\r^\dag}$ and ${b_\r = b_\r^\dag}$ satisfy ${\{a_\r, a_{\r'}\} = \{b_\r, b_{\r'}\} = 2\del_{\r,\r'}}$. In terms of these Majorana operators, the Hamiltonian is
\begin{gather}\label{tbHamReal}
    H = H_a + H_b,\\
    H_a = -\frac{i}2 \! \sum_{\<\rw,\rb\>} \!\!a_{\rw}a_{\rb},
    \qquad
    H_b = -\frac{i}2 \! \sum_{\<\rw,\rb\>}\!\! b_{\rw}b_{\rb}.\nonumber
\end{gather}
Interestingly, the two species of Majorana fermions $a_\r$ and $b_\r$ are completely decoupled. As we now show, an advantage of the Majorana fermions is that they make the existence of additional \textit{local} conserved charges of $H$ manifest. 

First, we note that the Hamiltonian~\eqref{tbHamReal} is independently invariant under lattice translations for each Majorana species. For instance, for each Bravais lattice vector $\x$, $H$ commutes with the $b$-Majorana translation operator
\begin{equation}
    T^{(b)}_{\x} \, a_\r \,(T^{(b)}_{\x})^{-1} = a_\r,
    \qquad 
    T^{(b)}_{\x}\,  b_\r \,(T^{(b)}_{\x})^{-1} = b_{\r+\x}.
\end{equation}
Therefore, by acting $T^{(b)}_{\x}$ on the conserved fermion number ${Q = \frac{i}{2}\sum_{\r\in\La} a_\r b_\r}$, we find a new, exactly conserved, integer quantized operator for each $\x$:\footnote{Each $Q_\x$ commutes with $H$ since $T^{(b)}_{\x}$ and $Q$ commute with $H$. Furthermore, they are integer quantized because $Q$ is integer quantized and $T^{(b)}_{\x}$ is a unitary operator. We can construct even more conserved charges using other $b$-Majorana crystalline symmetries. However, these charges would always be nonlocal.}
\ie \label{QxDef}
    Q_{\x} &=  T^{(b)}_{\x} Q \,  (T^{(b)}_{\x})^{-1}
    = \frac{i}{2}\sum_{\r\in\La} a_\r b_{\r+\x}.
\fe 
The corresponding current operator for this Hamiltonian is
\begin{equation}
    J^{(\x)}_{\langle \rw,\rb\rangle} = -\frac{i}{2}\left(a_{\rw} b_{\rb + \x} + a_{\rb} b_{\rw+\x}\right).
\end{equation}
Its lattice divergence ${(\mathrm{div}J^{(\x)})_{\r} = i\,[H, q^{(\x)}_\r]}$,\footnote{The lattice divergence is defined as ${(\mathrm{div}J^{(\x)})_{\rw} \equiv \sum_{\rb}J^{(\x)}_{\langle \rw,\rb \rangle}}$ for white sites and ${(\mathrm{div}J^{(\x)})_{\rb} \equiv -\sum_{\rw}J^{(\x)}_{\langle \rw,\rb \rangle}}$ for black sites.} where ${q^{(\x)}_\r = \frac{i}2 a_\r b_{\r+\x}}$ is the charge density of $Q_\x$ (i.e., ${Q_\x = \sum_{\r\in\La} q_\r^{(\x)}}$).
The unitary $e^{i\th Q_\x}$ implements the $\uone$ symmetry transformation 
\ie \label{eq:Qx-real-space}
    e^{i\th Q_\x} a_\r e^{-i\th Q_\x} &= a_{\r} \cos(\th) + b_{\r+\x} \sin(\th),\\
    e^{i\th Q_\x} b_\r e^{-i\th Q_\x} &=- a_{\r-\x} \sin(\th) + b_\r \cos(\th) .
\fe 

The charge operator $Q_\x$ in terms of the complex fermions~\eqref{comptoreal} is
\ie \label{eq:Qx-cplx}
Q_{\x} &= 
\frac{1}{2}\sum_{\r} \left(c_{\r} + (-1)^{\r} c_{\r}^\dagger\right)\left(c_{\r+\x} - (-1)^{\r} c_{\r+\x}^\dagger\right),
\fe 
where we define ${(-1)^{\r}=+/-}$ for ${\r \in \La_{\rm W/B} }$. Similarly, the current operator is
\begin{equation}
\begin{aligned}
    J^{(\x)}_{\langle \rw,\rb\rangle} = -\frac{i}2 \bigg(&(c_{\rw} + c_{\rw}^\da)(c_{\rb+\x} + c^\da_{\rb+\x}),\\
    &+(c_{\rb}-c^\da_{\rb})(c_{\rw+\x} - c^\da_{\rw+\x})
    \bigg).
\end{aligned}
\end{equation}
The U(1) symmetry transformation~\eqref{eq:Qx-real-space} in terms of the complex fermions is
\begin{equation}
\begin{aligned}
    e^{i\th Q_\x} c_\r e^{-i\th Q_\x} 
    &=\,
    c_\r \cos(\th) - \frac{i}2\bigg(c_{\r-\x}+ c_{\r+\x} \\
    &
    \hspace{12pt}+ (-1)^\r (c^\da_{\r-\x} - c^\da_{\r+\x})\bigg)\sin(\th).
\end{aligned}
\end{equation}

When the components of $\x$ are order one, the charge $Q_\x$ is local and $e^{i\th Q_\x}$ is a locality preserving unitary operator. The total symmetry formed by all $e^{i\th Q_\x}$ includes nonlocal unitary operators and, strictly speaking, does not correspond to an internal symmetry. This is similar to lattice translations. In both cases, the symmetry operators are not on-site, and while a generic symmetry operator is nonlocal, some of them are locality preserving.\footnote{Another similarity to lattice translations is that $e^{i\th Q_\x}$ are not on-site but compatible with a trivial gapped, unique ground state. For example, the Haldane mass term~\eqref{HaldaneTerm} commutes with all $e^{i\th Q_\x}$.}

The charges $Q_\x$ do not commute with each other. Instead, defining ${G_{\x} = \frac{i}2 \sum_{\r\in \La} (a_\r a_{\r+\x} - b_\r b_{\r+\x})}$, we find that they satisfy the non-Abelian algebra
\begin{equation}\label{2dOnsagerAlg}
\begin{aligned}
    [Q_{\x_1}, Q_{\x_2}] &= i G_{\x_2 - \x_1},
    \qquad
    [G_{\x_1},G_{\x_2}] = 0,\\
    [Q_{\x_1}, G_{\x_2}] &= 2i ( Q_{\x_1 -  \x_2}- Q_{\x_1 + \x_2} ).
\end{aligned}
\end{equation}
This algebra, which we denote by $\otwo$, can be generated by ${Q\equiv Q_{\mb{0}}}$, $Q_{\mb{a}_1}$, $Q_{\mb{a}_2}$, and $Q_{\mb{a}_1 + \mb{a}_2}$. It closely resembles the Onsager algebra~\cite{PhysRev.65.117}, which similarly appears in the ${1+1}$D staggered fermion model~\cite{CPS240912220}. In fact, $\otwo$ has infinitely many Onsager subalgebras when ${L_1,L_2\to\infty}$. For instance, there is an Onsager algebra for each primitive lattice vector $\mb{a}_i$, generated by ${Q}$ and $Q_{\mb{a}_i}$. The full algebra $\otwo$ is a type of generalized Onsager algebra defined by the group of lattice translations. In Appendix~\ref{OnsagerAppendix}, we further discuss this class of generalized Onsager algebras, their realization in fermionic lattice models, and the justification for the notation $\otwo$.

Unlike the original $\uone$ fermion number symmetry generated by $Q$, the Onsager symmetry generated by $Q_\x$ ($\x\neq\mb{0}$) transforms momentum space operators in a $\k$-dependent fashion. In particular, $e^{i \th Q_\x}$ acts differently on operators near the two Dirac cones (see Eq.~\eqref{gammaOnsActionApp}). As we will show later, these new symmetries of~\eqref{tbHam} become the SU(2) valley symmetry of the low-energy effective theory describing the two massless Dirac fermions. For this reason, we will often refer to $Q_\x$ as valley charges and the symmetries they generate as exact lattice valley symmetries.

\section{Symmetries and anomalies in the IR}

Before deriving this relation, let us first review relevant aspects of the continuum quantum field theory for the lattice Hamiltonian~\eqref{tbHam}. Its Lagrangian,\footnote{The complex 2-component field $\Psi$ is a Dirac field and $\Psi^\dag$ denotes its Hermitian conjugate (which is also a column vector). We use the Minkowski spacetime metric ${\eta_{\mu \nu} = \text{diag}(-1,1,1)}$ and choose gamma matrices ${\ga_0 = -i\si^y}$, ${\ga_1 = \si^z}$, and ${\ga_2 = \si^x}$, which satisfy the Clifford algebra ${\{\ga_\mu, \ga_\nu \} = 2 \eta_{\mu \nu}}$. By choosing real-valued gamma matrices, Majorana fermions are 2-component real fermions. As usual, the Dirac bar is defined as ${\bar{\Psi} \equiv (\Psi^\dag)^\mathrm{tr} \ga_0}$, where $\mathrm{tr}$ denotes transpose, and the slash notation ${\slashed{\pp} \equiv \ga_\mu\pp^\mu}$.}
\begin{equation}\label{continuum2DiracQFT}
    \mathscr{L} = i\bar{\Psi}^{\,(a)}\spp \Psi^{(a)} + i\bar{\Psi}^{\,(b)}\spp \Psi^{(b)}\,,
\end{equation}
has a manifest SU(2) flavor symmetry transforming the doublet ${(\Psi^{(a)}, \Psi^{(b)})^\top}$. The total internal symmetry group is O(4), which most naturally acts on the four Majorana fermions of the Dirac fields. Similarly, there is an ${\mathrm{O}(2)^a\times \mathrm{O}(2)^b\subset \mathrm{O}(4)}$ symmetry that implements independent O(2) transformations on $\Psi^{(a)}$ and $\Psi^{(b)}$. There are also charge conjugation, reflection, and time-reversal symmetries, and we denote their symmetry operators by $\cC$, $\cR$, and $\cT$, respectively. They satisfy ${\cC^2 = 1}$, $\cR^2=1$, and ${\cT^2 = (-1)^F}$, where the fermion parity ${(-1)^F}$ generates the center of SU(2), and act on the Dirac fermion fields as (we suppress the $(a), (b)$ superscripts)
\ie\label{contCRT}
   & \cC \Psi(t,x,y)  \cC^{-1} = \Psi^\dag(t,x,y),\\
& \cR \Psi(t,x,y) \cR^{-1} = \ga_1 \Psi(t,-x,y),\\
    &\cT \Psi (t,  x,y) \cT^{-1} = \ga_0 \Psi(-t, x,y).
\fe

The $\Z_4^{\cC\cT}$ symmetry generated by $\cC\cT$ has a ${4 \bmod 16}$ gravitational 't Hooft anomaly, while the $\Z_4^\cT$ symmetry generated by $\cT$  is free of anomalies~\cite{SW160204251}.\footnote{There is also a time-reversal symmetry operator $\t\cT$ that acts on the fermions as ${\t\cT \Psi^{(a)} (t,x,y)\t{\cT}^{-1}=- \gamma_0 \Psi^{(b)}(-t,x,y)}$ and ${\t{\cT} \Psi^{(b)} (t,x,y)\t{\cT}^{-1}=\gamma_0 \Psi^{(a)}(-t,x,y)}$. This $\Z_2^{\t\cT}$ symmetry is free of 't Hooft anomaly and commutes with the SU(2) flavor symmetry.}
There is also a parity anomaly, 
which is a mixed 't Hooft anomaly between SU(2) and any spacetime reflection (e.g., time-reversal $\cT$)~\cite{Niemi:1983rq, Redlich:1983kn, Redlich:1983dv, Alvarez-Gaume:1984zst}. It manifests in flat spacetime through the explicit breaking of spacetime reflections once a background SU(2) gauge field is turned on. 
More generally, the partition function $\cZ[A,g]$ of Eq.~\eqref{continuum2DiracQFT} with SU(2) background gauge field $A$ and spacetime metric $g$ transforms under a spacetime reflection as~\cite{Alvarez-Gaume:1984zst,Witten:2015aba,Witten:2016cio,SW160204251,Witten:2019bou}
\begin{equation}\label{eta}
     \cZ[A,g]\to e^{-i\pi  \eta[A,g]}\cZ[A,g]\,.
\end{equation}
Here, $\eta[A,g]$ is the $\eta$ invariant, which is defined as ${\lim_{\epsilon\to 0^+}\sum_i e^{- \epsilon |\lambda_i|}\,\text{sign}(\lambda_i)}$, where $\lambda_i$ are the eigenvalues of the ${2+1}$D Dirac operator coupled to $A$ and $g$ and ${\text{sign}(0)=1}$.
The ${3+1}$D anomaly inflow theory is  ${\exp(i \pi \int_{4d}\left( \frac{1}{96\pi^2}\text{tr}R\wedge R
-\frac{1}{8\pi^2}\text{tr}F\wedge F\right))}$, where $R$ is the Riemann tensor and $F$ is the field strength for SU(2).\footnote{The anomalous transformation~\eqref{eta} can be written in terms of the more familiar Chern-Simons functions as follows~\cite{Witten:2015aba,Witten:2016cio,SW160204251,Witten:2019bou}. We assume the Dirac fermions are in a representation $E$ of a global symmetry group $G$ (for~\eqref{continuum2DiracQFT}, ${G = \mathrm{SU}(2)}$ and $E$ is the fundamental rep). We extend the 3-dimensional spacetime spin manifold (and the various background fields) to  4 dimensions. Then, the Atiyah-Patodi-Singer theorem~\cite{Atiyah:1975jf} states that ${\mathfrak{J}=\int_{4d}\left( \frac{\text{dim}(E)}{ 192\pi^2}\text{tr}R\wedge R
-\frac{1}{ 8\pi^2}\text{tr}F\wedge F\right)  - \frac{\eta}{2}}$, where $\mathfrak{J}$ is the index (which is an integer) for the ${3+1}$D Dirac operator. This implies that ${\exp(-i \pi \eta) = \exp( i \text{CS}(A) 
 - 2i\text{dim}(E) \text{CS}_\text{grav})}$, where CS denotes the Chern-Simons functions. However, the overall phase of the 3D partition function ${\cZ[A] = |\cZ[A]| \exp(\mp i \pi \eta/2)}$~\cite{Alvarez-Gaume:1984zst} (where the signs are exchanged under spacetime reflection) cannot be rewritten in terms of the Chern-Simons functions because the latter are only well-defined modulo $2\pi \mathbb{Z}$.} This parity anomaly is generally a mod 2 anomaly on orientable spacetimes. See~\cite{Witten:2016cio, GPW171111587} for further discussion regarding
nonorientable spacetimes.

\section{Parity anomaly from Onsager symmetries}

We now return to the lattice model and relate its $\otwo$ symmetry charges to the symmetries of the continuum theory. In terms of the Majorana fermions, it is clear that the Hamiltonian ${H=H_a+H_b}$ in Eq.~\eqref{tbHamReal} factorizes into two decoupled systems, each of which is the Majorana honeycomb model studied by Kitaev~\cite{K0506438}. 
As reviewed in Appendix~\ref{App:MajHoneycombModel}, $H_b$ can be diagonalized in momentum space with dispersion
\begin{equation}\label{mainDisp}
    |1+ e^{i \k\cdot \mb{a}_1}+ e^{i \k\cdot \mb{a}_2}|.
\end{equation}
Even though this dispersion has two zeros at the ${\bf K}$ and ${\bf K}'$ points of the Brillouin zone, the momentum operators at these two points are related. 
Therefore, the IR limit of $H_b$ is a single massless Dirac fermion. 
This follows similarly for $H_a$. 

From here on, we let the continuum Dirac fermion $\Psi^{(a)}$ in the Lagrangian~\eqref{continuum2DiracQFT} be the one that arises from $H_a$, and $\Psi^{(b)}$ be the one from $H_b$. This basis for the two continuum Dirac fermions will be convenient for our analysis of the new exact valley symmetries below, but is different than the typical one corresponding to the Dirac cones at the $\mb{K}$ and $\mb{K}'$ points. Instead, each continuum fermion field is a superposition of the Dirac cone fields.

The fermion number $\uone$ symmetry charge~\eqref{FermionNum} becomes the charge $\cQ$ in the continuum, which  acts on $\Psi^{(a)}$ and $\Psi^{(b)}$ in the same way as $Q$ acts on $a_\r$ and $b_\r$:
\begin{equation}\label{cQ}
    e^{i \th \cQ}\, \binom{\Psi^{(a)}}{\Psi^{(b)}} \, e^{-i \th \cQ} = \begin{pmatrix}
        \cos(\th) & \sin(\th)\\
        -\sin(\th) & \cos(\th)
    \end{pmatrix}
    \binom{\Psi^{(a)}}{\Psi^{(b)}}.
\end{equation}
In other words, the continuum charge $\cQ$ is represented as the $\si^y$ matrix when acting on ${ (\Psi^{(a)},\Psi^{(b)})^\top}$. Therefore, the lattice $\uone$ fermion number symmetry becomes a $\uone$ subgroup of the SU(2) flavor symmetry. 

What about the $\otwo$ symmetry charges $Q_\x$? To find their continuum limit, which we denote as $\cQ_\x$, it suffices to find the continuum limit of $T^{(b)}_\x$ and then use Eq.~\eqref{QxDef}. When $\x$ is order 1, $T^{(b)}_\x$ becomes an internal symmetry of the continuum QFT.\footnote{It is common that lattice translation symmetry becomes an internal global symmetry in the continuum limit. For example, the lattice translation in the antiferromagnetic Ising model in ${1+1}$D with a small magnetic field becomes an internal $\mathbb{Z}_2$ symmetry in the IR.} Since it acts nontrivially only on $b_\r$, the corresponding continuum internal symmetry acts nontrivially only on the $\Psi^{(b)}$ continuum field and is part of O(2)$^b$. In Appendix~\ref{App:MajHoneycombModel}, we show that $ T^{(b)}_{\mb{a}_1}$ and $ T^{(b)}_{\mb{a}_2}$, respectively, become $e^{i \frac{2\pi}3 \cQ^{(b)}}$ and $e^{-i \frac{2\pi}3 \cQ^{(b)}}$ in the continuum limit, where $\cQ^{(b)}$ generates the $\uone$ symmetry
\begin{equation}\label{cQb}
    e^{i \th \cQ^{(b)}} \binom{\Psi^{(a)}}{\Psi^{(b)}}  \,  e^{-i \th \cQ^{(b)}} 
    =
    \begin{pmatrix}
        1 & 0\\
        0 & e^{-i\th}
    \end{pmatrix}
    \binom{\Psi^{(a)}}{\Psi^{(b)}}
    .
\end{equation}
The fact that $ T^{(b)}_{\mb{a}_1}$ and $ T^{(b)}_{\mb{a}_2}$ become order 3 elements of O(2)$^b$ in the continuum is consistent with the dispersion~\eqref{mainDisp} having exact zero modes only when $L_1$ and $L_2$ are both multiples of 3. Indeed, changing $L_i$ by one corresponds to inserting a defect for the order 3 internal global symmetry defect in the continuum.

\begin{figure}[t]
    \centering
    \begin{tikzpicture}[scale=2,>=Triangle, thick]
    \definecolor{aColor}{HTML}{d95f02}
        \draw[line width = 1.25pt] (0,0) -- (1,0) node[right]{$\cQ \equiv \cQ_{\mb{a}_1+\mb{a}_2}$};
        \draw[line width = 1.25pt] (0,0) -- (-0.72,-.215) node[above, yshift = 2pt]{$\cQ_{\mb{a}_1}$};
        \draw[line width = 1.25pt] (0,0) -- (-0.205,.295) node[above]{$\cQ_{\mb{a}_2}$};
        \draw (0,0) circle (1);
        \draw[dotted] (0,0) ellipse (1 and 0.3);
        \draw[-{>[scale=1.1]},line width = 1.25pt] (0,0) -- (0,1)  node[below right,xshift = 2pt, yshift = -1pt]{$\si^z$};
        \draw[-{>[scale=1.1]},line width = 1.25pt] (0,0) -- (1,0)  node[above left,xshift = 1pt]{$\si^y$};
        \draw[-{>[scale=1.1]},line width = 1.25pt] (0,0) -- (-0.707107/2.4,-.707107/2.4) node[yshift= 2pt, xshift = 6pt, below left]{$\si^x$};
        \draw[fill=white, draw=black] (1,0) circle (1.5pt);
        \draw[fill=white, draw=black] (-0.72,-.215) circle (1.5pt);
        \draw[fill=white, draw=black] (-0.205,.295) circle (1.5pt);
        \draw (-0.32,-0.1) to[out=-5,in=200] (0.6,0) node[yshift= -11pt, xshift = -30pt, right]{\footnotesize $2\pi/3$};
    \end{tikzpicture}
    \caption{The continuum limits of the on-site symmetry charge $Q$ and the valley charges $Q_\x$ generate an SU(2) internal global symmetry parametrized by a two-sphere.}
    \label{ContOnsagerFig}
\end{figure}
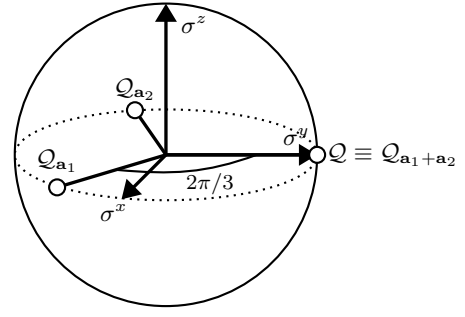

Now that we have understood the continuum limits of $Q$ and $T^{(b)}_{\mb{a}_i}$, we immediately find that the $\otwo$ charge $Q_\x$ (with $\x = n_1 \mb{a}_1 + n_2 \mb{a_2}$) in the continuum becomes
\begin{equation}\label{eq:Qx-su2}
\begin{aligned}
    \cQ_\x &= e^{i \frac{2\pi}3 (n_1 - n_2) \cQ^{(b)}} \cQ \, e^{-i \frac{2\pi}3 (n_1 - n_2) \cQ^{(b)}}.
\end{aligned}
\end{equation}
Using Eqs.~\eqref{cQ} and~\eqref{cQb}, it follows that the continuum charge $\cQ_\x$ acts on $(\Psi^{(a)},\Psi^{(b)})^\top$ as an SU(2) rotation represented by the matrix (see Fig.~\ref{ContOnsagerFig}) 
\begin{equation}
\begin{aligned}
 \sin\left(\frac{2\pi}3(n_1-n_2)\right)\si^x + \cos\left(\frac{2\pi}3(n_1-n_2)\right)\si^y
 \end{aligned}
\end{equation}
when acting on $(\Psi^{(a)},\Psi^{(b)})^\top$. 
Therefore, the generalized Onsager algebra $\otwo$ becomes the $\mathfrak{su}(2)$ Lie algebra of the SU(2) symmetry in the continuum.\footnote{In fact, the SU(2) valley symmetry arises from any Onsager subalgebra of $\otwo$ that is generated by $Q$ and $Q_{n_1\mb{a}_1+n_2\mb{a}_2}$ with ${n_1-n_2\neq0\bmod 3}$.} 
This SU(2) flavor symmetry is often referred to as a ``valley symmetry'' and viewed as an emergent symmetry of the continuum. Here, we find an exact avatar of this valley symmetry on the lattice, but it forms a much larger symmetry group than SU(2).\footnote{An essential aspect of this lattice valley symmetry is that it coexists with time-reversal and lattice reflection symmetries. It is straightforward to realize an exact lattice SU(2) symmetry by explicitly breaking spacetime reflection symmetries. For instance, consider two complex fermions ${(c_\uparrow, c_\downarrow)}$ per site governed by the tight-binding Hamiltonian $H$ of~\eqref{tbHam} plus the 
explicit $\sfT$-breaking deformation $\del H_{\text{Sem}}$ of~\eqref{SemenoffTerm}, i.e., $H_{\text{total}} = H_{\uparrow} + \del H_{\uparrow}+H_{\downarrow} + \del H_{\downarrow}$. For carefully chosen parameters $t'$ and $m$ in~\eqref{SemenoffTerm} and \eqref{HaldaneTerm}, $H_{\text{total}}$ can have two Dirac cones, one for each fermion species~\cite{H1988}. Then, there is an on-site SU(2) symmetry rotating the doublet ${(c_\uparrow,\, c_\downarrow)^\top}$ that becomes the SU(2) flavor symmetry of the continuum.} 
Even though the generators $Q$, $Q_{\mb{a}_1}$, $Q_{\mb{a}_2}$, and $Q_{\mb{a}_1+\mb{a}_2}$ of $\otwo$ are local, their nested commutators give more and more nonlocal charges. Therefore, it is not clear how to gauge this exact valley symmetry  on the lattice, despite the fact its IR limit is an SU(2) symmetry with no 't Hooft anomaly by itself.

The lattice time-reversal   and reflection symmetries become some spacetime orientation-reversing symmetries in the continuum. 
In Appendix~\ref{app:CRT}, we show that $\sfT$ becomes ${\cC\cT e^{-\frac{i\pi}{2}(\cQ^{(a)} -\cQ^{(b)})}}$ and $\sfR$ becomes $\cR\cC$ in the continuum limit. 
Therefore, the lattice model has symmetries that become the SU(2) and spacetime orientation-reversing symmetries in the continuum, which implies that the parity anomaly arises from the Onsager,  time-reversal, and lattice reflection symmetries.

\section{LSM anomaly of Onsager symmetries}\label{anomalyConsSect}

Since the symmetries for the parity anomaly in the IR come from exact lattice symmetries, it is natural to wonder whether there is an LSM anomaly for the Onsager, time-reversal, and lattice reflection symmetries that matches the parity anomaly. This LSM anomaly would mean that there is no trivial unique, gapped ground state in any lattice model with these symmetries. In fact, the SU(2) parity anomaly in the continuum is believed to force the low-energy phase to be either gapless or spontaneously break spacetime reflections~\cite{2014PhRvB..89s5124W,Dumitrescu:2024jko}, suggesting that the corresponding LSM anomaly is also incompatible with topological order. 
We now show that such an LSM anomaly exists in fermion lattice models.

What are deformations that preserve all of these symmetries? A straightforward generalization of the argument from~\cite{CPS240912220, PCS241218606} shows that the most general local Hamiltonian that has the Onsager symmetry generated by $Q_\x$ 
is ${\sum_{\y}(\t{g}^{(\mathrm{B})}_\y \t{H}^{(\mathrm{B})}_\y+\t{g}^{(\mathrm{W})}_\y \t{H}^{(\mathrm{W})}_\y + g_\y H_\y)}$, where ${\t{g}^{\,(\bullet)}_\y}$ and ${g_\y}$ are real-valued constants, $\y$ is a Bravais lattice vector whose components are order 1, and 
\begin{align}
\t{H}^{(\bullet)}_\y &= i\sum_{\r\in \La_{\bullet}}
\left( c^\dag_{\r}\, c_{\r +\y}
-c_{\r +\y}^\dag\, c_{\r}\right) \,,\label{dH2}\\
H_\y &= \sum_{\r\in \La_\mathrm{W}}
\left( c^\dag_\r \, c_{\r +\y +\bm{\del}}
+c^\dag_{\r +\y +\bm{\del}}\, c_{\r}\right)
\,.\label{dH0}
\end{align} 
Both terms in~\eqref{dH2} explicitly break the $\Z^\sfT_2$ symmetry, which enforces ${\t{g}^{\,(\mathrm{B})}_\y = \t{g}^{\,(\mathrm{W})}_\y = 0}$. On the other hand, the term~\eqref{dH0} will generally break some spatial reflections. For instance, the Hamiltonian $H_{\mb{a}_1-\mb{a}_2}$ commutes with each $Q_\x$ and $\sfT$, but breaks $\sfR$. This Hamiltonian is an array of decoupled ${1+1}$D gapped Majorana chains, which has a nondegenerate gapped ground state~\cite{Kitaev:2000nmw}.

The most general Hamiltonian  commuting with $Q_\x, \sfT$ and lattice reflections  is $H_\text{gen}={\sum_\y C_{|\y+\bm{\del}|} H_\y}$, 
where the constant ${C_{|\y+\bm{\del}|}}$ depends only on the magnitude ${|\y+\bm{\del}|}$.\footnote{This Hamiltonian has the entire crystalline symmetry --- requiring $Q_\x$ to commute automatically gives lattice translations while requiring lattice reflections automatically gives the rest of the crystalline symmetry group.}
The simplest  family of such  Hamiltonians   is
\begin{align}\label{HNNNN}
\sum_{\langle \r, \r'\rangle} c_\r^\dag c_{\r'}+
 t''\!\!\sum_{\langle\langle\langle \r, \r'\rangle\rangle\rangle} c^\dag_\r c_{\r'}\, .
\end{align}
As shown in Appendix~\ref{app:symdef}, this Hamiltonian is always gapless (see Fig.~\ref{fig:phase}). 
More generally, we show  in Appendix~\ref{gaplessApp} that the class of Hamiltonians $H_\text{gen}$ is always gapless and has two Dirac cones appearing at the $\mb{K}$ and $\mb{K}'$ points of the Brillouin zone. 
This is an LSM anomaly for the Onsager and spacetime reflection symmetries on the lattice.

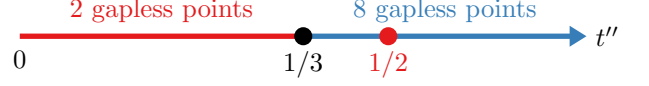
\begin{figure}[t]
    \centering
    \begin{tikzpicture}[scale=.75, >=Triangle, thick]
        \definecolor{niceRed}{HTML}{e41a1c}
        \definecolor{niceBlue}{HTML}{377eb8}
        \draw[niceBlue,-{>[scale=0.75]}, line width=2pt] (3,0) -- (8,0);
        \draw[niceRed, line width=2pt] (-2,0) -- (3,0);
        \node[black, right] at (8,0) {\normalsize $t''$};
        \filldraw (3,0) circle (4pt);
        \filldraw[niceRed] (4.5,0) circle (4pt);
        \node[niceRed, above, yshift=2pt] at (.5,0) {\normalsize 2 gapless points};
        \node[niceBlue, above, yshift=2pt] at (5.5,0) {\normalsize 8 gapless points};
        \node[black, below, yshift=-2pt] at (3,0) {\normalsize $1/3$};
        \node[niceRed, below, yshift=-2pt] at (4.5,0) {\normalsize $1/2$};
        \node[black, below, yshift=-2pt] at (-2,0) {\normalsize $0$};
       
    \end{tikzpicture}
    \caption{The phase diagram of~\eqref{HNNNN}, which preserves both the Onsager symmetry and spacetime reflections. The number of gapless modes changes from 2 to 8 at ${t'' = \frac 13}$ where there is a Lifshitz transition, and there is an additional point ${t''=\frac 12}$ where there are two gapless modes with quadratic dispersions.}
    \label{fig:phase}
\end{figure}

We now consider the fate of this LSM anomaly upon explicitly breaking the symmetries of~\eqref{tbHam}. In particular, consider the familiar deformations
\begin{align}
\delta H_{\text{Sem}}  &= 
 m \sum_{\rw}  c^\dag_{\rw} c_{\rw}
-m \sum_{\rb}  c^\dag_{\rb} c_{\rb},
\label{SemenoffTerm}
\\
\delta H_{\text{Hal}}  &= 
i t' \sum_{\langle\langle \r ,\r'\rangle\rangle} \epsilon_{\r\r'} c_\r^\dag c_{\r'}.\label{HaldaneTerm}
\end{align}
The first deformation $\delta H_{\text{Sem}}$ is a term considered by Semenoff that gaps out the Dirac cones and drives the system to a trivial, unique gapped ground state~\cite{S1984}. 
While $\delta H_{\text{Sem}}$ commutes with $\sfT$, it does not commute with some lattice reflections nor with $Q_{\x\neq\mb{0}}$. In the second deformation $\delta H_{\text{Hal}}$, ${\langle\langle \r ,\r'\rangle\rangle}$ denotes next-to-nearest neighbor pairs and $\epsilon_{\r\r'}$ is $+1$ if $\langle\langle \r ,\r'\rangle\rangle$ is pointing clockwise, and $-1$ otherwise (see Fig.~\ref{fig:honeycomb-lattice}). 
This is the Haldane mass term and it drives the system to a Chern insulator 
with a unique gapped ground state and no topological order~\cite{H1988}. $\delta H_{\text{Hal}}$ commutes with all $Q_\x$, but explicitly breaks the lattice reflection and time-reversal symmetries. Both deformations can drive the model to a trivial phase, which is consistent with the LSM anomaly since they explicitly break either the Onsager or spacetime reflection symmetries.

Another important property of the SU(2) parity anomaly is that it is a mod 2 anomaly on orientable spacetimes, so stacking the system with itself trivializes the anomaly. We can show this similarly applies for the LSM anomaly on the lattice. Stacking the system with itself amounts to first doubling the local degrees of freedom, so we now consider two flavors of complex fermions $c_{j,\si}$ per site $j$ differentiated by the ${\si = \uparrow, \downarrow}$ subscripts (i.e., they are spinful fermions). Then, consider the Hamiltonian
\begin{equation}
    \sum_{\langle \r,\r' \rangle, \si}  c_{\r\si}^\dag c_{\r'\si} 
    +
    \bar{t}\sum_{\langle\langle \r, \r' \rangle\rangle} \eps_{\r\r'} \left(c_{\r\uparrow}^\dag c_{\r'\downarrow} - c_{\r\downarrow}^\dag c_{\r'\uparrow} \right),
\end{equation}
where, as for the Haldane term~\eqref{HaldaneTerm}, ${\epsilon_{\r\r'} = +1}$ if hopping clockwise and ${\epsilon_{\r\r'} = -1}$ otherwise. This Hamiltonian commutes with quantized charges ${Q_\x^\uparrow + Q_\x^\downarrow}$ which generate the diagonal ${\otwo}$ subalgebra of ${\otwo\oplus \otwo}$. Furthermore, it commutes with the time-reversal symmetry acting as complex conjugation, and lattice reflection symmetries that also exchange the $\uparrow,\downarrow$ labels. It is quadratic and exactly solvable, from which we find that it has a unique gapped ground state. This is consistent with the LSM anomaly being order 2.

\section{Outlook}

In this paper, we have discussed a novel condensed matter realization of an SU(2) parity anomaly in ${2+1}$D. The key insight was the realization of an exact microscopic valley symmetry that became the SU(2) valley symmetry in the IR. This microscopic valley symmetry was an Onsager symmetry whose conserved, quantized symmetry charge operators obeyed a generalized Onsager algebra. We showed that it has an LSM anomaly with time-reversal and lattice reflection symmetries that becomes the parity anomaly in the IR. Similar anomaly-matching mechanisms have been found to capture other interesting 't Hooft anomalies on the lattice. This includes Schwinger's anomaly~\cite{Schwinger:1959xd}, which is the anomaly between U(1) vector and U(1) axial symmetries of a massless Dirac fermion in 1+1D, and was found to arise from an Onsager symmetry in Ref.~\onlinecite{CPS240912220}. It also includes Witten's anomaly for SU(2) global symmetry of two Weyl fermions in 3+1D~\cite{Witten:1982fp}, which was shown in Ref.~\onlinecite{GT250307708} to arise from an Onsager symmetry too.

The lattice realization of the parity anomaly discussed here is different than that of Ref.~\onlinecite{GT250307708}. There, a lattice $\R$ symmetry was constructed that becomes the $\uone$ symmetry of a single Dirac cone with parity anomaly. Unlike the exact lattice valley charges $Q_{\mb{a}_1}$ and $Q_{\mb{a}_2}$ constructed in this paper, the charge operator of this $\R$ symmetry was not quantized and its symmetry operator was not locality preserving.

There are many interesting follow-up questions. For example, Fermi surfaces in ${2+1}$D and higher are protected by an infinite-dimensional Lie group symmetry with an 't Hooft anomaly~\cite{Else:2020jln, Else:2025hxo}. It would be interesting to explore whether LSM anomalies of Onsager symmetries match these exotic 't Hooft anomalies. Dimensions ${2+1}$D and higher also support crystalline defects, like disclinations and dislocations. It would be interesting to investigate how inserting these defects modifies Onsager symmetries since, as we discussed, Onsager symmetries share many similarities with crystalline symmetries. Lastly, with now three established examples of 't Hooft anomalies arising from Onsager symmetries---Schwinger's anomaly, parity anomalies, and Witten's anomaly---it is tantalizing to wonder what is the more general relation between algebras and anomalies. For instance, what is the relation between these three 't Hooft anomalies that allow them to arise from Onsager symmetries? More broadly, one can ask which 't Hooft anomalies in quantum field theory can be matched by Onsager-type symmetries?

\let\oldaddcontentsline\addcontentsline
\renewcommand{\addcontentsline}[3]{}
\section*{Acknowledgments}
\let\addcontentsline\oldaddcontentsline

We would like to thank  Tom Banks, Maissam Barkeshli, Meng Cheng, Yichul Choi, Thomas Dumitrescu, Lukasz Fidkowski, Theo Jacobson, Patrick Ledwith, Chihiro Matsui, Nathan Seiberg, Senthil Todadri,  Wucheng Zhang, and Yunqin Zheng for interesting discussions.  
We are grateful to Nathan Seiberg and Wucheng Zhang for valuable discussions about their project~\cite{SZ}. 
We further thank Maissam Barkeshli, Zohar Komargodski,  Nathan Seiberg, and Wucheng Zhang for comments on the draft, and express our special gratitude to Maissam Barkeshli and Meng Cheng for  discussions which partially inspired this project. 
SDP is supported by the NSF GRF under Grant No. 2141064 and by grant NSF PHY-2309135 to the Kavli Institute for Theoretical Physics (KITP).
MLK and AC are supported by NSF DMR-2022428 and by the Simons Collaboration on Ultra-Quantum Matter, which is a grant from the Simons Foundation (651446, Wen).
SHS is also supported by
the Simons Collaboration on Ultra-Quantum Matter 
(651444, Shao). 
 Part of this work
was completed during the KITP program 
``Generalized Symmetries in Quantum Field Theory: High Energy Physics, Condensed Matter, and Quantum Gravity,'' 
which is supported in part by grant NSF PHY-2309135 to the KITP.

\appendix

\section{Crystalline symmetries of the honeycomb lattice} \label{app:crystalSymsHoneycomb}

Here, we briefly review the crystalline symmetries of the honeycomb lattice. When the honeycomb lattice is placed on a torus---when it has periodic boundary conditions---the crystalline symmetries form the group ${D_{12}\ltimes (\Z_{L_1}\times \Z_{L_2})}$. Its point group is the order 12 Dihedral group ${D_{12}\cong \Z_6\rtimes \Z_2}$, which can be generated by the plaquette-centered ${2\pi/6}$ rotation $P$, with its center of rotation at ${-\bm\del=(0,1)}$, and the reflection $\rfl$ that transforms ${(x,y)\to(-x,y)}$. The translation group ${\Z_{L_1}\times \Z_{L_2}}$ can be generated by the lattice translations $T_{\mb{a}_1}$ and $T_{\mb{a}_2}$ by the primitive lattice vectors $\mb{a}_1$ and $\mb{a}_2$, respectively (see Fig.~\ref{fig:honeycomb-lattice}). These generators satisfy the ${D_{12}\ltimes (\Z_{L_1}\times \Z_{L_2})}$ relations
\begin{equation}\label{HoneyCombCryAlg}
    \begin{aligned}
    &\rfl^2  = P^6 = T_{\mb{a}_1}^{L_1}  =  T_{\mb{a}_2}^{L_2}  =1,\quad   T_{\mb{a}_1} T_{\mb{a}_2}=T_{\mb{a}_2} T_{\mb{a}_1}\,, \\
    &\rfl\,  P \, \rfl= P^{-1} , \quad   \rfl\,  T_{\mb{a}_1} \rfl=T_{\mb{a}_2}, \quad   \rfl\,  T_{\mb{a}_2} \rfl=T_{\mb{a}_1}\,,\\  
    &P \, T_{\mb{a}_1} P^{-1} =T_{\mb{a}_2}, \quad  
    P \, T_{\mb{a}_2} P^{-1}=T_{\mb{a}_1}^{-1}\, T_{\mb{a}_2} \,.
    \end{aligned}
\end{equation}
When the honeycomb lattice is placed on $\R^2$, its translation group becomes $\Z^2$, and its crystalline symmetry group becomes the wallpaper group ${p6m\cong D_{12}\ltimes \Z^2}$. 

Throughout this paper's main text and appendices, we often use some elements of this crystalline symmetry group more than others. These include the generators $T_{\mb{a}_1}$ and $T_{\mb{a}_2}$ of the translation group, the reflection $\rfl$, and two 3-fold site-centered rotations: one centered at ${(0,0)\in \La_{\rm W}}$ called $U_{\rm W}$ and another centered at ${\bm\del\in\La_{\rm B}}$ called $U_{\rm B}$. In terms of the above generators of the crystalline symmetry group,
\ie \label{UWandUBdef}
U_{\rm W} = T_{\mb a_1}^{-1} P^2 \,,\quad 
U_{\rm B} = T_{\mb a_1}^{-2} P^2
\,.
\fe 

Let us discuss the action of these symmetries on the lattice more explicitly. We will use the notation ${g \cdot \r}$ to denote the lattice site obtained from the action of the crystalline symmetry $g$ on the site $\r\in\La$. Translations $T_{\mb a_i}$ act as ${T_{\mb a_i}\cdot \r = \r + \mb a_i}$. The rotation $U_{\rm W}$ acts as
${U_{\rm W}\cdot (\mb a_1, \mb a_2, \bm \del) 
= (\mb a_2 - \mb a_1, - \mb a_1, \mb a_1 + \bm \del )}$. 
The reflection $\rfl$ acts as
${\rfl\cdot (\mb a_1, \mb a_2, \bm \del) 
= (\mb a_2, \mb a_1, \bm \del)}$.
The action of $U_{\rm W}$ and $\rfl$ on arbitrary ${\r\in\La}$ can be obtained by extending the above actions linearly. 
The action of the rotation $U_{\rm B}$ follows from the relation ${U_{\rm B} = T_{\mb a_1}^{-1} U_{\rm W}}$ (cf.~\eqref{UWandUBdef}).

\section{\texorpdfstring{$G$-Onsager algebras in a general tight-binding model}{G-Onsager algebras in a general tight-binding model}}\label{OnsagerAppendix}

In this Appendix, we present a generalization of the Onsager algebra based on a group $G$ and discuss its physical realization in lattice fermions. Various generalizations of the Onsager algebra exist in the math and physics literature.\footnote{See~\cite{UI9502068, MR2056679, BK0507053, BB09061215, BS09061482, S181007408, M220316594} for an (incomplete) survey of various generalizations of the Onsager algebra.} We will call the generalization presented here the $G$-Onsager algebra. $G$ can be a discrete or continuous group, and the $\Z$-Onsager algebra is isomorphic to Onsager's original algebra~\cite{PhysRev.65.117}.

The $G$-Onsager algebra is a Lie algebra with elements denoted by $Q_g$ and $G_{g,h}$, where ${g,h\in G}$. The $G$-Onsager algebra relations are
\ie \label{eq:G-algebra}
    [Q_g, Q_h] &= 
        i G_{g^{-1}h,\, hg^{-1}}, \\
    [G_{g_1,h_1},G_{g_2,h_2}] &= i ( G_{g_2g_1,\,h_1h_2} - G_{g_1g_2,\,h_2h_1}\\
    &\hspace{20pt}+ G_{g_1g_2^{-1},\,h_2^{-1}h_1} - G_{g_2^{-1}g_1,\,h_1h_2^{-1}} ),\\
    [Q_g, G_{h,k}] &= i \left( Q_{gh^{-1}} + Q_{k^{-1}g} - Q_{gh}  - Q_{kg} \right).
\fe 
It is straightforward to confirm that the Jacobi identity is satisfied. Furthermore, from the first algebraic relation, the element ${G_{g^{-1}h,\, hg^{-1}} = - G_{h^{-1}g,\, gh^{-1}}}$, which implies that ${G_{g,\, hgh^{-1}}=0}$ for all ${h\in G}$ if ${g^2 = 1}$. Therefore, the $G$-Onsager algebra is non-Abelian if there exists a ${g\in G}$ such that ${g^2 \neq 1}$. 

For an Abelian subgroup ${A\leq G}$, the elements $Q_a$ and ${G_{a,a}\equiv G_a}$, with ${a\in A}$, form the $A$-Onsager subalgebra
\begin{equation}\label{AbelianSubAlg}
\begin{aligned}
    [Q_{a_1}, Q_{a_2}] &= i G_{a_2 a_1^{-1}},\\ 
    [G_{a_1},G_{a_2}] &= 0,\\
    [Q_{a_1}, G_{a_2}] &= 2i ( Q_{a_1 a_2^{-1}}- Q_{a_1 a_2} ).
\end{aligned}
\end{equation}
Note that ${G_a = 0}$ if ${a^2 = 1}$. When ${A = \Z_2}$, all ${G_a=0}$, and this is the Lie algebra ${\mathfrak{u}(1)\oplus \mathfrak{u}(1)}$ generated by $Q_1$ and $Q_{-1}$. When ${A= \Z}$, this is the Onsager algebra, which can be generated by $Q_0$ and $Q_{1}$. When ${A=\Z^n}$, we refer to the $A$-Onsager algebra as the degree-$n$ Onsager algebra, denoted by $\on$.

The $G$-Onsager algebra is naturally represented by ${N_f = 2}$ Majorana fermion operators on a spatial lattice $\La$ with crystalline symmetry $G$. Let $a_\r$ and $b_\r$ denote the Majorana operators at site ${\r\in \La}$, and ${g\cdot \r}$ the action of ${g\in G}$ on the lattice vector $\r$. The $G$-Onsager algebra elements are then represented by the operators
\begin{align}
    Q_g &= \frac{i}{2}\sum_{\r\in \La} a_\r b_{g \,\cdot \r},\label{generalQgOnsagerCharge}\\
    G_{g,\, h} &= \frac{i}2 \sum_{\r\in \La} (a_\r a_{g\,\cdot \r} - b_\r b_{h\,\cdot \r}).\label{generalGgOnsagerCharge} 
\end{align}
Indeed, from the anti-commutation relations of $a_\r$ and $b_\r$, it is straightforward to show that the operators~\eqref{generalQgOnsagerCharge} and~\eqref{generalGgOnsagerCharge} satisfy~\eqref{eq:G-algebra} with the Lie bracket being the commutator. When $G$ is the one-dimensional translation group $\Z$, this representation of the Onsager algebra was discussed in~\cite{CPS240912220}. In the main text, these operators were discussed for the two-dimensional translation group ${\Z\times\Z}$.

If the lattice $\La$ is $d$-dimensional bipartite, there is a simple lattice model Hamiltonian that commutes with the $\Z^d$-Onsager charges~\eqref{generalQgOnsagerCharge}. Denoting by $\rw$ and $\rb$ the sublattice sites forming the bipartite structure, the Hamiltonian is
\begin{equation}\label{genTBHam}
    H = \sum_{\<\r,\r'\>}c_\r^\da c_\r = -\frac{i}2 \! \sum_{\<\rw,\rb\>} \!\! (a_{\rw}a_{\rb}+b_{\rw}b_{\rb}).
\end{equation}
The complex fermions $c_\r$ are related to the real fermions $a_\r$ and $b_\r$ as~\eqref{comptoreal} in the main text.
Following the reasoning in the main text, this is a tight-binding model with vanishing chemical potential, and there is an integer quantized, conserved charge ${Q = \frac{i}2\sum_{\r\in\La} a_\r b_\r}$. The Hamiltonian~\eqref{genTBHam} also commutes with the $b$-Majorana translation operator $T_{\x}^{(b)}$ by the lattice vector ${\x\in\La\cong \Z^d}$. The Onsager charge $Q_\x$ can be written as ${Q_\x = T_\x^{(b)} Q (T_\x^{(b)})^{-1}}$. Since $Q$ and $T_\x^{(b)}$ commute with $H$, acting the unitary operator $T_\x^{(b)}$ on $Q$ constructs an integer quantized, conserved operator. Therefore, the Onsager charge ${Q_\x}$ is an integer quantized, conserved operator of the Hamiltonian~\eqref{genTBHam}.

\subsection{Square lattice}

A simple ${d=2}$ bipartite lattice is an ${L_x\times L_y}$ square lattice with both $L_x$ and $L_y$ an even integer. The Hamiltonian~\eqref{genTBHam} for the square lattice is a tight-binding model with dispersion ${2(\cos(k_x) + \cos(k_y))}$. This model has a Fermi surface, which forms a square in momentum space, and its low energy modes depend on an angle denoting their location on the Fermi surface. The Majorana translation $T^{(b)}_\x$ and, consequentially, the Onsager symmetry transformation $e^{i\th Q_\x}$ also depend on this angle. We leave for future work elucidating how the Onsager symmetry transformation embeds into the LU(1) symmetry that protects this Fermi surface~\cite{Else:2020jln}.

\section{Majorana honeycomb model}\label{App:MajHoneycombModel}

In this Appendix, we discuss the Majorana honeycomb lattice model of~\cite{K0506438} and its continuum limit, which is a single, free, massless Dirac fermion. 
This is "half" of the model~\eqref{tbHam} considered in the main text.

We derive the symmetry operator algebra, which is generally realized projectively on the Hilbert space, by analyzing the fermion zero modes. 
We follow the ${1+1}$D analysis in~\cite{Seiberg:2023cdc} closely.

\subsection{The Hamiltonian}
\label{app:maj-ham-diag}

There is a single Majorana fermion $b_\r$ at every site of the honeycomb lattice, and the
Hamiltonian is given by (cf.~\eqref{tbHamReal})
\ie\label{eq:Hb}
H_b = -\frac{i}2 \! \sum_{\<\rw,\rb\>}\!\! b_{\rw}b_{\rb}\,.
\fe
This Hamiltonian can be diagonalized by passing to momentum space, defining ${b_{\rm W} (\k) = \frac{1}{\sqrt{N}}\sum_{\rw} e^{-i\k \cdot \rw} b_{\rw}}$ and ${b_{\rm B} (\k) = \frac{1} {\sqrt{N}}\sum_{\rb} e^{-i\k \cdot (\rb-\bm{\del})} b_{\rb}}$, where the number of unit cells (each containing two sites) in the lattice is denoted by ${N = L_1 L_2}$. These momentum modes satisfy
\ie \label{eq:bk-conj}
& \hspace{13pt} b_{\rm W}^\dag(\k)  = b_{\rm W}( -\k),
\qquad 
b_{\rm B}^\dag(\k) = b_{\rm B}(-\k)\,,\\
& \{b_{\rm W}(\k) , b_{\rm W}(\k') \} = 
\{b_{\rm B}(\k) , b_{\rm B}(\k') \} = 2 \delta_{\k,- \k'}\,.
\fe 
The diagonalized form of $H_b$ is 
\ie \label{eq:Hbdiag}
H_b = \frac12 \sum_{\k \in \bz} |\Delta_{\k}| \, \beta^\dag_\k \beta_\k\,,
\fe 
where $\bz$ is the first Brillouin zone, which is defined as the Wigner-Seitz unit cell of the reciprocal lattice, and
\begin{equation}\label{Disp}
     \Del_\k = 
    1+ e^{i \k\cdot \mb{a}_1}+ e^{i \k\cdot \mb{a}_2}.
\end{equation}
We have also defined the new momentum modes, ${\beta_\k = \frac{1}{\sqrt 2}\left[ i e^{- i\th_{\k}} b_{\rm W}(\k) + b_{\rm B}(\k) \right]}$, where ${\th_{\k} = \arg(\Delta_{\k}) \in (-\pi , \pi]}$.\footnote{Note that this definition is only sensible when $\Delta_\k\neq 0$. However, $\k$ points with ${\Delta_\k = 0}$ do not contribute to the Hamiltonian~\eqref{eq:Hbdiag}, so this doesn't affect the diagonalization.\label{fn:Delta0}}
The anti-commutation relations of the new momentum modes $\beta_\k$ are
\ie \label{eq:beta-acr}
\{\beta_{\k},\beta_{\k'}\} = 0\, , \quad  \{\beta_{\k},\beta_{\k'}^\dagger\} = 2\delta_{\k,\k'}\,.
\fe 
The reciprocal lattice is spanned by ${\mb{b}_1 = \frac{2\pi}{3} \left(\sqrt{3}, 1 \right)}$ and $ {\mb{b}_2 = \frac{2\pi}{3} \left(-\sqrt{3}, 1 \right)}$.
From~\eqref{eq:Hbdiag}, 
the low-energy states are located near ${\mb{K} = \frac{1}{3} \mb{b}_1 + \frac{2}{3} \mb{b}_2}$ and ${\mb{K'}=\frac{2}{3} \mb{b}_1 + \frac{1}{3} \mb{b}_2}$, where ${\Del_\k = 0}$. 
Expanding the dispersion ${|\Del_\k|}$ about ${\k\approx \mb K,\mb K'}$, we find that the low-energy excitations satisfy ${E \sim |\k-\mb K^\bullet|}$. 
Therefore, the effective IR description is that of a single, free, massless Dirac fermion $\Psi$, which is denoted  as $\Psi^{(b)}$ in the main text.

For the purposes of identifying a dictionary between the symmetries of the Majorana honeycomb model and the IR Dirac fermion field theory, it is sufficient to restrict to their respective zero modes. 
The Hamiltonian $H_b$ has  4 exact zero modes   when ${L_1,L_2 =0 \bmod 3}$ with periodic boundary conditions, so we will assume that for the rest of this Appendix.

\subsection{Lattice zero modes}

The lattice zero modes are spanned by $b_{\bullet}(\mb K)$ and their adjoints, where $\bullet$ stands for either of the two sublattices, B/W. Using~\eqref{eq:bk-conj}, we can express every operator at $\mb K'$ in terms of those at $\mb K$. 
We decompose the four independent zero mode operators into real fermions
\ie
    \mu_\bullet &= \frac1{\sqrt{2}}(b_\bullet(\mb{K})+b^\dagger_\bullet(\mb{K})),\\
    \mu'_\bullet &= \frac1{i\sqrt{2}}(b_\bullet(\mb{K})-b^\dagger_\bullet(\mb{K})).
\fe
These real fermions form a ${\rm Cl}(4,0)$ Clifford algebra. 
We choose the following ${4\times4}$ matrix representations for these zero modes,
\ie \label{eq:mu-defs-1flavor}
&\mu_{\rm B} = \si^z\otimes\one\,,
\qquad  
\mu_{\rm W} = \si^x\otimes\si^x\,,
\\ 
&\mu_{\rm B}' = \si^y\otimes\one\,,
\qquad  
\mu_{\rm W}' = \si^x\otimes\si^z\,.
\fe

The action of the lattice symmetry operators on these zero modes can be deduced from their action on the lattice Majorana operators.
These actions are implemented by unitary operators
${O_g \, b_\r \, O_g^{-1} = b_{g \cdot \r} }$.
Here, ${g \cdot \r}$ denotes the action of the crystalline symmetry $g$ on lattice site $\r$, as in Appendix~\ref{app:crystalSymsHoneycomb}.\footnote{We commit a slight abuse of notation throughout the paper, often writing the unitary operator $O_g$ simply as $g$.}
For instance, the lattice translation $T_{\mb a_i}$ acts as ${T_{\mb a_i} b_\r T_{\mb a_i}^{-1}=b_{\r + \mb a_i}}$.\footnote{In the main text, these symmetries are referred to as $T^{(b)}_{\mb a_i}$. In this Appendix, we will drop the superscript $(b)$ for the lattice symmetry operators since we only consider the $b$ Majorana fermion.}

We also consider  two (anti-unitary) time-reversal symmetry operators:
\ie 
\trs_{\rm W} b_{\rw} \trs_{\rm W}^{-1} &= -b_{\rw}\,, \quad  \trs_{\rm W} b_{\rb} \trs_{\rm W}^{-1} = b_{\rb}\,, \\
\trs_{\rm B} b_{\rw} \trs_{\rm B}^{-1} &= b_{\rw}\,, \quad  \hspace{10.5pt} \trs_{\rm B} b_{\rb} \trs_{\rm B}^{-1} = -b_{\rb}\,. 
\fe 
They differ by the fermion parity $(-1)^F$ symmetry, which flips the sign of every $b_\r$. 
Both of these two time-reversal symmetries satisfy $(\sfT_\bullet)^2=1$, i.e., each generates a $\mathbb{Z}_2^\sfT$ symmetry. We do not find a $\mathbb{Z}_4^\sfT$ symmetry in this Majorana lattice model, i.e., there is no lattice time-reversal symmetry that squares to $(-1)^F$.

We find the explicit matrix representations for the above lattice symmetries acting on the ground space spanned by the zero modes~\eqref{eq:mu-defs-1flavor}:
\ie\label{eq:latalg}
&T_{\mb a_1}=e^{-\frac{i\pi}{3}\si^x}\otimes e^{-\frac{i\pi}{3}\si^y}\,,~~~
&&T_{\mb a_2}= e^{\frac{i\pi}{3}\si^x}\otimes e^{\frac{i\pi}{3}\si^y}\,,\\
&U_{\rm W} = - (e^{ -\frac{i\pi}{3}\sigma^x}\otimes \mathbf{1} ) \,,~~~
&&
U_{\rm B} =-( \mathbf{1}\otimes e^{ \frac{i\pi}{3}\sigma^y} ) \,,\\
&\sfR = \sigma^z\otimes \sigma^z\,,~~~
&&(-1)^F = \sigma^x\otimes \sigma^y\,,\\
&\sfT_{\rm W} = (\sigma^z\otimes \mathbf{1})\cK\,,~~~
&&\sfT_{\rm B}= (\sigma^y\otimes \sigma^y)\cK\,,
\fe
where $\cK$ denotes the complex conjugation operator. 
For any state $\ket{\psi}$ in 
the 4-dimensional ground space spanned by the zero modes, they satisfy 
\ie \label{eq:lattice-1maj-translation}
&(T_{\mb a_1})^3\ket{\psi}=(T_{\mb a_2})^3\ket{\psi} = T_{\mb a_1} T_{\mb a_2} \ket{\psi}= \ket{\psi}\,,\\
&U_{\bullet} T_{\mb a_i}\ket{\psi} = T_{\mb a_i}U_{\bullet} \ket{\psi}\,,
\fe

From~\eqref{eq:latalg}, we derive the following  relations between the various symmetry operators that hold true on the entire Hilbert space: 
\ie\label{eq:lattice-1maj-symalg}
& (U_{\bullet})^3 =  \rfl^2 = (\trs_{\bullet})^2 = 1 \,,\\
&\sfR T_{\mb a_1} = T_{\mb a_2} \sfR\,,\
\trs_{\bullet}T_{\mb a_i}  = T_{\mb a_i} \trs_{\bullet} 
\,,\\
&  \sfR U_{\bullet} = U_{\bullet}^{-1} \sfR\,,\
U_\bullet \trs_{\bullet'} = \trs_{\bullet'}U_\bullet\,,\\
&\rfl \, \trs_{\bullet} = \trs_{\bullet} \rfl\,,
\fe 
where $\bullet$ stands for either B or W and $(-1)^F$ commutes with every operator.

\subsection{Continuum zero modes}\label{app:1Dirac-zeromodes}

Following \cite{Delmastro:2021xox}, we discuss the zero modes of the continuum field theory of a massless Dirac fermion field. 
We assume the space is a two-torus with periodic boundary conditions in both directions. 
We decompose the zero modes two complex components as
\begin{equation}\label{eq:Psi-decomp}
\Psi =\frac{1}{\sqrt{2}}
    \begin{pmatrix}
        \lambda_1 + i \lambda'_1 \\
        \lambda_2 + i \lambda'_2
    \end{pmatrix}.
\end{equation}
Using real gamma matrices, namely
${\gamma_0 = -i\si^y}$, ${\gamma_1 = \si^z}$, ${\gamma_2 = \si^x}$,
allows us to treat $(\la_1,\la_2)^\top$ and $(\la_1',\la_2')^\top$ as two Majorana fermions, each with two real components. 
 We choose the following matrix representation for these 4 real, zero modes:
\ie \label{eq:lambda-defs-1flavor}
&\la_1 = \si^z\otimes\one\,,\qquad  
\la_2 = \si^x\otimes\si^x\,,\\ 
&\la'_1 = \si^y\otimes\one\,,\qquad  
\la'_2 = \si^x\otimes\si^z\,.
\fe

Next, we discuss the global symmetry of a single Dirac fermion field theory. 
The internal global symmetry is O(2) generated by the charge $\cQ$ and charge conjugation operator $\cC$, which act as
${e^{i\th \cQ} \Psi e^{-i\th \cQ} = e^{-i\th} \Psi}$ and ${\cC \Psi \cC^{-1} = \Psi^\dag}$.\footnote{In the main text this symmetry and the charge operator are denoted as O(2)$^b$ and $\cQ^{(b)}$. Similar to the discussion of the lattice zero modes, in this Appendix we will drop the superscript $b$ for the symmetry groups and operators since here we focus on a single Dirac fermion field.}
Time-reversal and reflection symmetries act as ${\cT \Psi \cT^{-1} = \ga_0 \Psi}$ and ${\cR \Psi \cR^{-1} = \ga_1 \Psi}$. Finally, spatial rotation acts as ${\cU_\th \Psi \cU_\th^{-1} = e^{-i\theta \sigma^y/2} \Psi}$.
We find the explicit matrix representations for all of these symmetry operators acting on the 4-dimensional ground space  spanned by  zero modes in~\eqref{eq:lambda-defs-1flavor}:
\ie
&e^{i \theta\cQ} = e^{ \frac{i\th}{2}\si^x} \otimes e^{\frac{i\th}{2}\si^y}\,,~~~\cC=\si^z\otimes\si^z\,,~~~
\cR= \mathbf{1}\otimes \sigma^y\,\\
&\cT = \frac1{\sqrt{2}}\begin{pmatrix}
    0 & -i & -i & 0 \\
    1 & 0 & 0 & 1 \\
    -1 & 0 & 0 & 1 \\
    0 & i & -i & 0
\end{pmatrix} \cK\,,\\
&\cU_\th = 
\begin{pmatrix}
e^{-\frac{i\th}{4}} \cos\frac{\th}4 & 0 & 0 & e^{-\frac{i\th}{4}} \sin\frac{\th}4 
\\ 
0 & e^{\frac{i\th}{4}}\cos\frac{\th}4 & e^{\frac{i\th}{4}}\sin\frac{\th}4 & 0 
\\ 
0 & -e^{\frac{i\th}{4}}\sin\frac{\th}4 & e^{\frac{i\th}{4}}\cos\frac{\th}4 & 0 
\\ 
-e^{-\frac{i\th}{4}} \sin\frac{\th}4 & 0 & 0 & e^{-\frac{i\th}{4}} \cos\frac{\th}4
\end{pmatrix}.
\fe  
From this matrix representation, we find the following algebra realized on the Hilbert space~\cite{SZ}:\footnote{We thank Nathan Seiberg and Wucheng Zhang for  discussions on this point and for sharing results from their upcoming paper~\cite{SZ}. }
\ie \label{eq:1-dirac-zeromode-symmalg}
    &
    \cU_{4\pi} = e^{ 2\pi i \cQ} = \cC^2 =\cR^2 = 1\,,\
    \cU_{2\pi}=   \cT^2 = - e^{i\, \pi \cQ}\,, \\
    & \cC e^{i\th \cQ} =  e^{-i\th \cQ} \cC\,,\ 
    \cR e^{i\th \cQ} = e^{i\th \cQ} \cR\,,\ 
    \cT e^{i\th \cQ} =  e^{-i\th \cQ} \cT\,,  \\
    & \cC\cU_\theta = 
     \cU_\theta\cC \,,\ 
     \cR\cU_\theta = 
     \cU_\theta^{-1}\cR\,,\
     \cT\cU_\theta = 
     \cU_\theta\cT \,,
     \\
    & \cT \cR = - e^{i\, \pi \cQ} \cR \cT \,, \ 
    \cC \cR = - \cR \cC\,, \  
    \cC\cT = - \cT\cC\,,\\
    &e^{i\th'\cQ} \cU_\th = \cU_\th e^{i\th'\cQ}\,.
\fe 
In particular, the projective sign in ${\cC\cT = - \cT\cC}$ cannot be removed by phase redefinitions of these operators, indicating a mixed anomaly between $\Z_2^{\cC}$ and $\Z_4^{\cT}$.\footnote{The sign in $\cU_{2\pi }=-e^{i\pi\cQ}$ can be removed by redefining the rotation operator as $\tilde{\cU_\th}=e^{\frac{i\th}2} \cU_\th$ so that $\tilde\cU_{2\pi}=e^{i\pi\cQ}$. But this introduces phases in $\cR \tilde{\cU_\th}= e^{i\th}{\tilde \cU_\th}^{-1}\cR$ and $\cT \tilde{\cU_\th}= e^{- i \th} \tilde{\cU_\th}\cT$. This corresponds to redefining the lattice rotation operators as $\tilde U_{\rm W} =e^{\frac{i \pi}3} (e^{- \frac{i\pi}3 \si^x}\otimes \mathbf{1})$ and $\tilde U_{\rm B} =e^{\frac{i \pi} 3} (\mathbf{1}\otimes e^{ \frac{i\pi}3\si^y})$, which introduce similar phases in the lattice algebra.} There is a similar anomaly between $\Z_2^\cC$ and $\Z_2^\cR$ since $\cC\cR = - \cR \cC$. We also note that
\ie
\cT^2 = -(\cC\cT)^2 = \cU_{2\pi} =  - e^{i \pi \cQ},
\fe
where $e^{i\pi \cQ}$  is the fermion parity operator that flips the sign of $\Psi$.

\begin{table}
\centering
\begin{tabular}{c|c|c}
\toprule
    Symmetry & Lattice & Continuum \\[2pt]
    \hline
    Translation & $(T_{\mb a_1}, T_{\mb a_2})$ & $(e^{i \frac{2\pi }{3}\cQ}, e^{-i\frac{2\pi }{3} \cQ})$ \Tstrut
    \\[5pt]
    $\frac{2\pi}{3}$ Rotation & $(U_{\rm W}, U_{\rm B})$ & $(-\cU_{\frac{2\pi}{3}} e^{i\frac{\pi}{3} \cQ}, -\cU_{\frac{2\pi}{3}} e^{-i\frac{\pi}{3} \cQ})$  \\[5pt]
    Reflection & $\rfl$ & $i\cR\cC$ \\[5pt]
    Time Reversal & $(\trs_{\rm W},\trs_{\rm B})$ & $(\cC \cT e^{-i\frac{\pi}{2} \cQ}, \cC \cT e^{i\frac{\pi}{2} \cQ} )$ \\[5pt]
    Fermion parity & $(-1)^F$ & $e^{i\pi \cQ}$ 
    \\
    \bottomrule
\end{tabular}
\caption{Symmetry operators of the Majorana honeycomb lattice model, and their corresponding symmetry operators in the continuum. (We suppress the factors $e^{2\pi i \frac{\mb{a}_i\cdot \cP}{L_i}}$  in this table.)}
\label{tab:1maj-honeycomb-symm-match}
\end{table}

\subsection{Matching the continuum and lattice symmetries} \label{app:MajHoneycombSymmMatch}

With all of these pieces of the puzzle in place, we are ready to match the symmetries of the continuum field theory with 
those in the lattice model. 
The dictionary between lattice and continuum symmetries is shown in Table~\ref{tab:1maj-honeycomb-symm-match}. Below, we go through various examples and discuss the derivation.

The lattice model~\eqref{eq:Hb} has relatively few internal symmetries compared to the internal O(2) symmetry in the continuum, but it has time-reversal and reflection symmetries. This is to be contrasted with Haldane's lattice model~\cite{H1988}, another microscopic realization of a single massless Dirac fermion. While Haldane's model has an internal O(2) symmetry on the lattice, it does not have time-reversal and lattice reflection symmetries.

While most elements of the O(2) symmetry are only emergent in the IR of~\eqref{eq:Hb}, 
certain discrete elements still arise from the crystalline symmetries of the lattice model. 
For instance, two order 3 elements of O(2) arise from lattice translations 
$T_{\mb a_1}$ and $T_{\mb a_2}$:
\ie 
    T_{\mb{a}_1} = e^{i \frac{2\pi}3 \cQ} e^{2\pi i \frac{\mb{a}_1\cdot \cP}{L_1}},
    \quad 
    T_{\mb{a}_2} = e^{-i \frac{2\pi}3 \cQ} e^{2\pi i \frac{\mb{a}_2\cdot \cP}{L_2}},
\fe
where $\cP$ is the spatial momentum operator for the  translation symmetry in the continuum.   
This can be seen from the lattice algebra in~\eqref{eq:lattice-1maj-translation} on the ground space, which implies that these lattice translations flow to two order 3 internal symmetries in the IR that are inverses of each other.
Such IR symmetries, which arise from exact conserved lattice operators obeying different algebra are referred to as "emanant" symmetries in~\cite{ CS221112543,Seiberg:2023cdc,Barkeshli:2025cjs}.

The lattice rotation also mixes with the internal global symmetry in the continuum. 
Indeed, on the one hand, the lattice $2\pi/3$ spatial rotation $U_\bullet$ is an order 3 operator. On  the other hand, the continuum $2\pi/3$ spatial rotation $\cU_{2\pi/3}$ is an order 6 operator, satisfying $(\cU_{2\pi/3})^3 = - e^{i\pi \cQ}$, which is the fermion parity. 
A detailed zero mode analysis shows that the lattice rotation $U_\bullet$ is identified with the continuum symmetry $-\cU_{2\pi/3}e^{\pm i \pi\cQ/3}$.  
(See~\cite{Barkeshli:2025cjs} for related discussions for lattice models of complex fermions.)

The lattice time-reversal and reflection symmetries are also related to the continuum symmetries in a nontrivial way. Indeed, since   $\trs_\bullet$ square to $+1$ on the lattice, whereas the  continuum time-reversals $\cT$ and $\cC\cT$ do not, the continuum limit of $\trs_\bullet$ necessarily mixes 
with certain internal symmetry elements in the IR.

Let us briefly summarize our derivation of Table~\ref{tab:1maj-honeycomb-symm-match}.
The action of a lattice symmetry $\sf S$ on the lattice zero modes (collectively denoted as a 4-vector $\bm \mu$) is given by ${ {\sf S} \, {\bm \mu}\,  {\sf S}^{-1} = D( {\sf S}) \, {\bm \mu}}$, where $D({\sf S})$ is a ${4\times4}$ matrix. We would like to identify ${\sf S}$ with a continuum symmetry ${\cal S}$, acting on the continuum zero modes (collectively denoted as a 4-vector $\bm \la$) as ${{\cal S} \, {\bm \la}\, {\cal S}^{-1} = D({\cal S}) \, {\bm \la}}$.
To that end, we want to find a linear combination of the lattice zero modes, ${\t {\bm \lambda} = M {\bm \mu}}$ with $M$ being a ${4\times4}$ matrix, 
such that $\t{\bm \la}$ transforms like ${\bm \la}$ when acted on by ${\sf S}$, i.e.,
\ie \label{eq:sfS-la}
{\sf S}\, \t{\bm \la} \, {\sf S}^{-1} = D(\cal S) \t{\bm \la} \, .
\fe 
Therefore, $M$ has to obey
\ie \label{eq:tM-match-S}
M \, D({\sf S}) \,  M^{-1} = D(\cal S) \,.
\fe 
Furthermore, requiring that the components of $\t {\bm \la}$ are self-adjoint and satisfy the ${\rm Cl}(4,0)$ Clifford algebra forces $M$ to be a real orthogonal matrix. 
As suggested by the notation, we can then identify $\t{\bm \la}$ with the continuum zero modes, $\bm \la$, and the lattice symmetry ${\sf S}$ with the continuum symmetry $\cal S$.
(If $\sf S$ is an anti-unitary symmetry, we need to perform complex conjugation on $M$ in going from~\eqref{eq:sfS-la} to~\eqref{eq:tM-match-S}. However, since $M$ is real, this is a trivial operation and~\eqref{eq:tM-match-S} remains unchanged.)

As a nontrivial consistency check, one can verify that the lattice symmetry algebras in~\eqref{eq:lattice-1maj-translation} and~\eqref{eq:lattice-1maj-symalg} matches with the continuum algebra in  
\eqref{eq:1-dirac-zeromode-symmalg} under the correspondence in  Table~\ref{tab:1maj-honeycomb-symm-match}.

\section{More on the complex honeycomb model}

In this Appendix, we elaborate on the tight-binding honeycomb model of complex fermions in~\eqref{tbHam} and its IR limit of two free massless Dirac fermion fields in~\eqref{continuum2DiracQFT}. 
We present the explicit action of the exact lattice valley symmetries, generated by $Q_\x$, on the low-energy degrees of freedom. 
Then, we discuss some of the  symmetries of~\eqref{tbHam}, and identify how they are matched with the symmetries of the IR field theory, closely following the analogous treatment of the Majorana honeycomb model in Appendix~\ref{App:MajHoneycombModel}. 
Lastly, we consider deformations of the Hamiltonian~\eqref{tbHam} that preserve the Onsager, time-reversal, 
and lattice reflection symmetries. 
After exploring the phase diagram of the simplest family of symmetric models, we give a  proof that the most general local Hamiltonian respecting these symmetries is always gapless.

\subsection{The Hamiltonian}

In order to diagonalize the tight-binding Hamiltonian \eqref{tbHam}, we perform a Fourier transform as usual. We define the momentum modes as\footnote{These are related to $b_{\bullet}(\k)$ defined below \eqref{eq:Hb} and the analogously defined $a_\bullet(\k)$ by ${c_{\rm W}(\k) = \frac12 \left(a_{\rm W}(\k) + i b_{\rm W}(\k)\right)}$ and ${c_{\rm B}(\k) = \frac12 \left(b_{\rm B}(\k) - i a_{\rm B}(\k)\right)}$.}
\ie 
    c_\mathrm{W}(\k) &= \frac{1}{\sqrt{N}} \sum_{\rw \in \La_{\rm W}} e^{-i\k \cdot \rw} c_{\rw}\,,\\
    c_{\mathrm{B}}(\k) &= \frac{1}{\sqrt{N}} \sum_{\rb \in \La_{\rm B}} e^{-i\k \cdot (\rb-\bm\del)} c_{\rb}\,,
\fe 
where ${N\equiv L_1L_2}$ is the number of unit cells and $\bullet$ stands for either of the two sublattices, B/W. They satisfy the canonical anticommutation relations, namely, ${\{c_{\rm B}(\k),c^\dagger_{\rm B}(\k')\} = \del_{\k,\k'}}$, ${\{c_{\rm W}(\k),c^\dagger_{\rm W}(\k')\} = \del_{\k,\k'}}$, and all the other anticommutators vanish.

In terms of these momentum modes, the Hamiltonian becomes
\ie
H = \sum_{\k\in\bz}\begin{pmatrix} 
    c_{\rm W}^\dagger(\k) & c_{\rm B}^\dagger(\k)
    \end{pmatrix}
    \begin{pmatrix} 
    0 & \Delta_{\k} \\ \Delta_{\k}^* & 0
    \end{pmatrix}
    \begin{pmatrix} 
    c_{\rm W}(\k) \\ c_{\rm B}(\k)
    \end{pmatrix}
\fe 
where $\Del_\k$ is given in \eqref{Disp}.
To complete the diagonalization of $H$, we make a further change of variables for the nonzero modes (i.e., $\k \neq {\bf K}$ or ${\bf K}'$) to\footnote{The momentum modes $\ga_\k$ and $\tilde\ga_\k$ defined here are related to the $\bt_\k$ defined below~\eqref{eq:Hbdiag} and the analogously defined ${\al_\k = \frac{1}{\sqrt 2}\left[ i e^{- i\th_{\k}} a_{\rm W}(\k) + a_{\rm B}(\k) \right]}$ by the relations ${\gamma_\k = \frac{1}{2}(\al_\k + i \bt_\k)}$ and ${\tilde{\gamma}_\k = \frac{1}{2}(\al_\k - i \bt_\k)}$. }
\ie 
\ga_\k &= \frac{i}{\sqrt 2}\left[e^{-i\th_\k} c_{\rm W}(\k) + c_{\rm B}(\k) \right],
\\
\tilde\ga_\k &= \frac{i}{\sqrt 2}\left[ e^{- i\th_{\k}} c_{\rm W}^\dagger(-\k) - c_{\rm B}^\dagger(-\k)\right],
\fe 
where ${\th_\k=\arg\Del_\k\in(-\pi,\pi]}$.
They satisfy canonical anticommutation relations, similar to those of $c_\bullet(\k)$, i.e., $\{\gamma_\k, \gamma^\dag_{\k'}\} = 
    \{\tilde{\gamma}_\k, \tilde{\gamma}^\dag_{\k'}\} = \del_{\k,\k'}$.
In terms of these new variables, the Hamiltonian becomes (up to an unimportant additive constant),
\ie
H = \sum_{\k \in \bz} |\Delta_{\k}|
    \left(
    \gamma^\dag_\k \gamma_\k + \tilde{\gamma}^\dag_\k \tilde{\gamma}_\k 
    \right).
\fe

The low-energy states of $H$ lie near the $\mb K$ and $\mb K'$ points where ${\Del_\k=0}$. Expanding the dispersion $|\Del_\k|$ near these points, and restricting our attention to $\k = \q+\mb K^{\bullet}$ with $|\q|\approx 0$, we define
\ie 
\gamma_{\mb K^{\bullet}}(\q)=\gamma_{\mb K^{\bullet} + \q} \,,
\quad \tilde\gamma_{\mb K^{\bullet}}(\q)=\tilde\gamma_{\mb K^{\bullet} + \q} \,.
\fe

\subsection{Exact lattice valley symmetry action}\label{app:valley}

The on-site $\uone$ symmetry generated by the charge $Q= \sum_{\r\in \La} (c_\r^\dag c_\r - 1/2)$  acts on these momentum modes  as
\begin{align}\label{eq:U1Q-cd}
    e^{i \theta Q} \gamma_\k e^{-i \theta Q} = e^{-i\th} \gamma_\k\,,\quad
    e^{i \theta Q} \tilde{\gamma}_\k e^{-i \theta Q} = e^{i\th} \tilde{\gamma}_\k.
\end{align}
On the other hand, the lattice valley symmetries $e^{i\theta Q_{\mb a_i}}$ (defined in~\eqref{QxDef}) act on these operators as:
\begin{align}
    e^{i \theta Q_{\mb a_i}} \gamma_\k e^{-i \theta Q_{\mb a_i}} &=  (\cos(\th)-i\cos(\k\cdot\mb{a}_i)\sin(\th) ) \gamma_\k 
     \nonumber\\
    & \qquad 
    - \sin(\k\cdot\mb{a}_i)\sin(\th) \tilde\gamma_\k\,,\\
    e^{i \theta Q_{\mb a_i}} \tilde\gamma_\k e^{-i \theta Q_{\mb a_i}} &=  (\cos(\th)+i\cos(\k\cdot\mb{a}_i)\sin(\th) ) \tilde\gamma_\k 
     \nonumber\\
    & \qquad  
    + \sin(\k\cdot\mb{a}_i)\sin(\th) \gamma_\k \,.
\end{align}

For the low-energy degrees of freedom, near the $\mb{K}$ and $\mb{K}'$ points, these equations simplify to the following:
\ie \label{gammaOnsActionApp}
    e^{i\th Q_{\mb a_i}}
    \begin{pmatrix}
        \gamma_{\mb K}(\q)\\
        \tilde\gamma_{\mb K}(\q)
    \end{pmatrix} 
    e^{-i\th Q_{\mb a_i}}
    &= e^{i\th \mb n_{\mb a_i}\cdot\bm\si}
    \begin{pmatrix}
        \gamma_{\mb K}(\q)\\
        \tilde\gamma_{\mb K}(\q)
    \end{pmatrix},
    \\
    e^{i\th Q_{\mb a_i}}
    \begin{pmatrix}
        \gamma_{\mb K'}(\q)\\
        \tilde\gamma_{\mb K'}(\q)
    \end{pmatrix}
    e^{-i\th Q_{\mb a_i}}
    &=
    e^{-i\th \mb n_{\mb a_i}\cdot\bm\si}
    \begin{pmatrix}
        \gamma_{\mb K'}(\q)\\
        \tilde\gamma_{\mb K'}(\q)
    \end{pmatrix},
\fe
where ${\mb n_{\mb a_1} = -\frac{\sqrt{3}}2 \hat{\mb y} + \frac12 \hat{\mb z}}$ and ${\mb n_{\mb a_2} = - \mb n_{\mb a_1} }$.
The above equation explicitly demonstrates that  $e^{i \theta Q_{\mb a_i}}$ act on the two valleys differently. In contrast, the on-site symmetry $e^{i \th Q}$ acts on these low-energy degrees of freedom in a valley-independent manner,
\ie \label{eq:gammaQaction}
e^{i\th Q}
    \begin{pmatrix}
        \gamma_{\mb K}(\q)\\
        \tilde\gamma_{\mb K}(\q)
    \end{pmatrix} 
    e^{-i\th Q}
    &= e^{-i\th \si^z}
    \begin{pmatrix}
        \gamma_{\mb K}(\q)\\
        \tilde\gamma_{\mb K}(\q)
    \end{pmatrix},
    \\
    e^{i\th Q}
    \begin{pmatrix}
        \gamma_{\mb K'}(\q)\\
        \tilde\gamma_{\mb K'}(\q)
    \end{pmatrix}
    e^{-i\th Q}
    &=
    e^{-i\th \si^z}
    \begin{pmatrix}
        \gamma_{\mb K'}(\q)\\
        \tilde\gamma_{\mb K'}(\q)
    \end{pmatrix}.
\fe

\subsection{Matching 
 the continuum and lattice symmetries}\label{app:CRT}

The Hamiltonian~\eqref{tbHam} has 8 exact zero modes when ${L_1,L_2=0\bmod 3}$. 
This 16-dimensional ground space is spanned by $c_\bullet(\mb K)$, $c_\bullet(\mb K')$ and their adjoints. 
We decompose these into eight real fermions as
\ie 
\mu_\bullet(\mb K) &= \frac1{\sqrt{2}} (c_\bullet(\mb K) + c^\dagger_\bullet(\mb K))\,,\\
\mu'_\bullet(\mb K) &= \frac1{i\sqrt{2}} (c_\bullet(\mb K) - c^\dagger_\bullet(\mb K))\,,\\
\mu_\bullet(\mb K') &= \frac1{\sqrt{2}} (c_\bullet(\mb K') + c^\dagger_\bullet(\mb K'))\,,\\
\mu'_\bullet(\mb K') &= \frac1{i\sqrt{2}} (c_\bullet(\mb K') - c^\dagger_\bullet(\mb K'))\,.
\fe 
These eight real fermions satisfy the ${\rm Cl}(8,0)$ Clifford algebra. To analyze various discrete lattice symmetries, we choose a basis of real 16-dimensional representation for these operators.

The various crystalline and internal symmetries of~\eqref{tbHam} can be studied within the ground space. We will restrict our attention to a subset of these symmetries.
In particular, we pick matrix representations for the lattice charge conjugation $\sfC$, reflection $\sfR$, and time-reversal $\sfT$, whose actions on the  Majorana operators in position space are defined in~\eqref{latCRT}.\footnote{Among the various crystalline symmetries not considered in this Appendix are the 3-fold site-centered rotations, $U_{\rm W/B}$. These were discussed recently in~\cite{Barkeshli:2025cjs}.} We do the same for the on-site $\uone$ symmetry $e^{i\th Q}$. 
From these explicit matrix representations of the symmetries, we derive the following relations between them, which hold on the entire Hilbert space:
\ie \label{eq:lat-cpx-symalg}
    &
    \sfC^2 = \rfl^2 = \trs^2 = 1\,,\
    \sfC \trs = \trs \sfC\,, \  \sfC \rfl = \rfl \sfC\,,\ \trs \rfl = \rfl \trs\\  
    &  \trs e^{i\theta Q} = e^{-i\theta Q} \trs \,,\ 
    \sfC e^{i\theta Q} = e^{-i\theta Q} \sfC\,, \  \rfl e^{i\theta Q} = e^{i\theta Q} \rfl \,.
\fe 
The behavior of the lattice translations $T_{\mb a_1}$ and $T_{\mb a_2}$ in the zero mode subspace is similar to that in the Majorana honeycomb model discussed in~\eqref{eq:lattice-1maj-translation}. For any state $\ket\psi$ in the 16-dimensional ground space spanned by the zero modes, they satisfy
\ie 
(T_{\mb a_1})^3\ket{\psi}=(T_{\mb a_2})^3\ket{\psi} = T_{\mb a_1} T_{\mb a_2} \ket{\psi}= \ket{\psi} \,.
\fe

Let us relate this to the model's continuum limit, which is described by two free massless Dirac fermion fields  $\Psi^{(a)}$ and $\Psi^{(b)}$. The IR field theory has an ${\rm O}(4)$ internal symmetry, which contains an SU(2) subgroup that rotates the doublet ${(\Psi^{(a)},\Psi^{(b)})^\top}$.
There is also an  ${\rm O}(2)^a \times{\rm O}(2)^b$ subgroup of O(4) with each ${\rm O}(2)$ factor acting on $\Psi^{(a)}$ and $\Psi^{(b)}$ separately. 
The respective $\uone$ subgroups are generated by the charges $\cQ^{(a)}$ and $\cQ^{(b)}$, and we denote
the associated charge conjugation operators by
$\cC^{(a)}$ and $\cC^{(b)}$ respectively. 
 On the other hand, the on-site $\uone$ lattice symmetry flows to yet another $\uone$ in the continuum, whose generator is denoted by $\cQ$. 
These three $\uone$ subgroups of O(4) act on the Dirac fermion doublet as
\ie 
e^{i \th \cQ^{(a)}} \binom{\Psi^{(a)}}{\Psi^{(b)}}  \,  e^{-i \th \cQ^{(a)}} 
    &=
    \begin{pmatrix}
        e^{-i\th} & 0\\
       0&1
    \end{pmatrix}
    \binom{ \Psi^{(a)}}{\Psi^{(b)}}
    ,\\  
e^{i \th \cQ^{(b)}} \binom{\Psi^{(a)}}{\Psi^{(b)}}  \,  e^{-i \th \cQ^{(b)}} 
    &=\begin{pmatrix}
       1 & 0\\
       0& e^{-i\th}
    \end{pmatrix}
   \binom{\Psi^{(a)}}{ \Psi^{(b)}}
    ,\\  
 e^{i \th \cQ}\, \binom{\Psi^{(a)}}{\Psi^{(b)}} \, e^{-i \th \cQ} &= \begin{pmatrix}
        \cos(\th) & \sin(\th)\\
        -\sin(\th) & \cos(\th)
    \end{pmatrix}
    \binom{\Psi^{(a)}}{\Psi^{(b)}}.
    \fe
The lattice valley charges $Q_\x$ flow to the conserved operators $\cQ_\x$ in the continuum in \eqref{eq:Qx-su2}, which acts on the doublet as
\ie
 &e^{i \th \cQ_{\x}}\, \binom{\Psi^{(a)}}{\Psi^{(b)}} \, e^{-i \th \cQ_{\x}} \\
 &= \begin{pmatrix}
        \cos(\th) & \sin(\th)e^{\frac{2 \pi i(n_1-n_2)}3 }\\
        -\sin(\th) e^{\frac{-2 \pi i(n_1-n_2)}3 } & \cos(\th)
    \end{pmatrix}
    \binom{\Psi^{(a)}}{\Psi^{(b)}},\\
\fe
where ${\x=n_1 {\mb a}_1 +n_2{\mb a}_2}$. 
Note that while the $\uone$ symmetries generated by $\cQ$ and $\cQ_{\x}$ are  subgroups of ${\rm SU}(2)$, those generated by $\cQ^{(a)}$ and $\cQ^{(b)}$ are not elements of
${\rm SU}(2)$.

We perform the zero mode analysis for the continuum field theory on a spatial two-torus with periodic boundary conditions. The zero modes of each of the two Dirac fermions comprise four real components each (as reviewed in Appendix \ref{app:1Dirac-zeromodes}). 
As with the lattice zero modes, they form the ${\rm Cl}(8,0)$ Clifford algebra. We pick an explicit basis for these zero modes, as well as the symmetries of fermion parity $(-1)^F$, which is equal to ${e^{i \pi (\cQ^{(a)}+\cQ^{(b)})}=e^{i \pi \cQ}}$, charge conjugation $\cC$, reflection $\cR$, and time-reversal $\cT$. The spacetime symmetries, $\cR$ and $\cT$ act on each Dirac fermion as ${\cT \Psi \cT^{-1} = \ga_0 \Psi}$ and ${\cR \Psi \cR^{-1} = \ga_1 \Psi}$, where we have suppressed the superscripts $a,b$ and the spacetime coordinates.
There are various other symmetries of the field theory that we do not consider here, such as spatial rotation and general elements of the ${\rm O}(4)$ group.

We find the following relations between the different discrete symmetry operators
\ie \label{eq:cont-cpx-symalg}
&\cR^2 = \cC^2 =  1\,,\  
\cT^2  = e^{i \pi \cQ} = (-1)^F \,,
\\ 
& \cT \cR = (-1)^F \cR \cT\,, 
\  \cT \cC = \cC \cT\,, \ \cR \cC = \cC \cR \,,
\\
& e^{i\th\cQ} \cC = \cC e^{i\th\cQ}\,,
\ 
e^{i\th\cQ} \cR = \cR e^{i\th\cQ}\,,
\ e^{i\th\cQ} \cT = \cT e^{i\th\cQ}
\,.
\fe
In particular, the charge conjugation $\cC$, defined in~\eqref{contCRT}, is  ${\cC = \cC^{(a)}\cC^{(b)}}$, and commutes with $\cQ$.

We now consider the IR fate of the lattice symmetries of~\eqref{tbHam} discussed above. 
Since this lattice model is equivalent to two copies of the Majorana honeycomb model in Appendix~\ref{App:MajHoneycombModel}, formed by $a_j$ and $b_j$ fermions, we can use the map in Table~\ref{tab:1maj-honeycomb-symm-match} to find the relations between these symmetries in the UV and IR. We verify these by explicitly mapping their actions on the zero modes, similar to the analysis in Appendix~\ref{app:MajHoneycombSymmMatch}. 
For instance, we find that the lattice translations ${T_{\mb a_1}}$ and ${T_{\mb a_2}}$ mix with the order-3 elements of the flavor ${\rm SU}(2)$ symmetry in the IR, $e^{\pm \frac{2\pi i (\cQ^{(a)}+\cQ^{(b)})}{3}}$.
Similarly, the relation between the lattice and continuum time-reversal symmetries also involve other internal global symmetries. This is to be expected since the lattice time-reversal symmetry obeys ${\sfT^2=1}$, while the continuum $\cT$ (or $\cC \cT$) squares to $(-1)^F$.
We present a partial dictionary between the lattice and continuum symmetries in Table~\ref{tab:2maj-honeycomb-symm-match}.
The relations in~\eqref{eq:lat-cpx-symalg} and~\eqref{eq:cont-cpx-symalg} are consistent with each other under this dictionary.

\begin{table}
\centering
\begin{tabular}{c|c|c}
\toprule
    Symmetry & Lattice & Continuum \\[2pt]
    \hline
    Charge conjugation & $\sfC$ & $e^{i\pi\cQ^{(b)}} $  
    \Tstrut\\[5pt]
    Reflection & $\rfl$ & $\cR\cC$ \\[2.5pt]
    Time Reversal & $\trs$ & $\cC \cT e^{-\frac{i\pi}{2} (\cQ^{(a)}-\cQ^{(b)})} $
    \\
    \bottomrule
\end{tabular}
\caption{Symmetry operators of the lattice model~\eqref{tbHam}, and their corresponding symmetry operators in the continuum field theory of two free massless Dirac fermions. 
}
\label{tab:2maj-honeycomb-symm-match}
\end{table}

\subsection{Simplest symmetric deformation}\label{app:symdef}

Here, we discuss in detail the simplest deformation of the tight-binding Hamiltonian~\eqref{tbHam} which preserves the generalized Onsager algebra $\otwo$, time-reversal $\sfT$, and all the reflection symmetries. It is a next-to-next-to nearest neighbor (NNNN) Hamiltonian~\eqref{HNNNN} mentioned in the main text.

\begin{figure*}
    \centering
    \includegraphics[width=\linewidth]{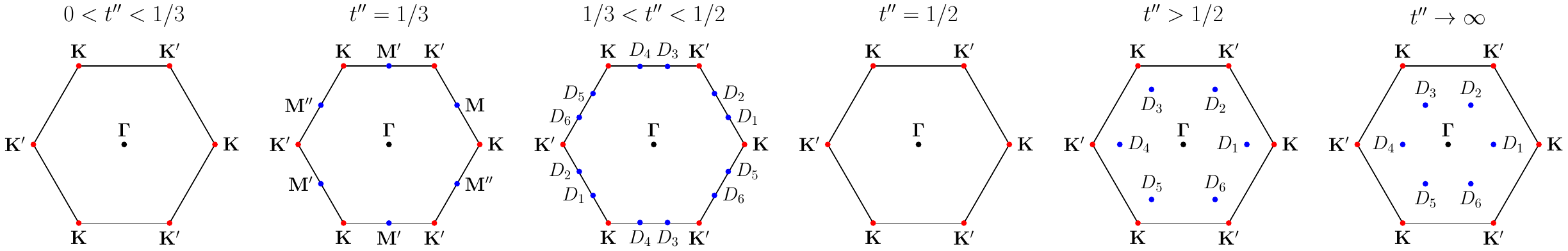}
    \caption{Locations of the gapless modes in the Brillouin zone of the Hamiltonian~\eqref{HNNNN} as $t''$ varies. For ${0 \leq t''<1/3}$, all gapless modes reside at the $\mb K/\mb K'$ points. Additional modes appear when ${t''=1/3}$ and moves throughout the Brillouin zone as $t''$ increases.}
    \label{fig:smallt}
\end{figure*}

Each lattice site has three NNNN sites. The NNNN sites of the lattice site at the origin are at ${\bm{\xi}_1 = (0,2)}$, ${\bm{\xi}_2 = (\sqrt{3}, -1)}$, and ${\bm{\xi}_3 =  (\sqrt{3}, 1)}$. Together with the nearest-neighbor term, the total Hamiltonian in momentum space is
\begin{equation}
    H = \sum_{\k \in \text{BZ}} (\Delta_{\k} + t'' \Xi_{\k}) c_{\rm W}^\dagger(\k) c_{\rm B}(\k) + \text{h.c.}.
\end{equation}
Here, $\Delta_\k$ is defined in \eqref{Disp} and $\Xi_{\k} = \sum_{n=1}^3 e^{i\k \cdot (\bm{\xi}_n - \bm{\delta})} = e^{3ik_y} + 2 \cos (\sqrt{3}k_x)$. There are two bands whose dispersions as a function of $t''$ are
\begin{equation}
    \pm|\Delta_{\k} + t''\Xi_{\k} |.
\end{equation}

This dispersion vanishes at the $\mb K$ and $\mb K'$ points, and there are Dirac cones there for all values of $t''$. Therefore, the total Hamiltonian is gapless for all $t''$, consistent with the LSM anomaly discussed in the main text. The number of gapless degrees of freedom, however, will vary with $t''$. For convenience, we will restrict our discussion to ${t'' \geq 0}$. Solving the gaplessness condition ${\Delta_{\k} + t''\Xi_{\k} = 0}$, we find how the number of gapless modes change as a function of $t''$ (see Fig.~\ref{fig:smallt}).

As $t''$ increases from zero, the first new gapless modes appear when ${t''=1/3}$, and arise at the ${\mb M = (\frac{\pi}{\sqrt{3}}, \frac{\pi}{3})}$, ${\mb M' = (0, \frac{2\pi}{3})}$, and ${\mb M'' = (-\frac{\pi}{\sqrt{3}}, \frac{\pi}{3})}$ points of the Brillouin zone.\footnote{The additional gapless modes are not isotropic Dirac cones, and instead elliptical cones. For example, the gapless mode at ${\k = (\frac{\pi}{\sqrt{3}}, 0)}$ when ${t''=1}$ has the dispersion relation ${\sqrt{3k_x^2 + 9k_y^2}}$.} This leads to a total of 5 gapless points occurring at ${t''=1/3}$. As $t''$ is further increased, each of the 3 gapless points at $\mb M$, $\mb M'$, and $\mb M''$ split into two, causing the model to enter a gapless phase where the total number of gapless modes is 8. As further $t''$ increases, these modes move towards $\mb K$ and $\mb K'$ points, respectively, until ${t''=1/2}$ when they collide with the Dirac cones $\mb K$ and $\mb K'$ points and the number of gapless modes at this point specifically becomes 2 again. However, as ${t''>1/2}$, these gapless modes reappear and travel along the line connecting $\Gamma$ and $\mb K/\mb K'$ points as $t''$ increases. As $t''$ tends to infinity, they approach the midpoint between the $\Gamma$ and $\mb K/\mb K'$ points, which is the purely NNNN hopping result. Therefore, as shown in Fig.~\ref{fig:phase}, there are two phase transitions occurring at ${t'' = 1/3}$ and ${t'' = 1/2}$.

The dynamical exponents of these transitions are easily extracted from the dispersion relation at the phase transition points. For example, at the ${t''=\frac{1}{3}}$ transition, the $\mb M/\mb M'/\mb M''$ gapless modes' dispersions are not isotropic. Their dispersion is quadratic when moving alongside the BZ edge but linear in the orthogonal direction. Therefore, there are two dynamical critical exponents ${z_\| = 2}$ and ${z_\perp = 1}$ at ${t''=\frac{1}{3}}$. On the other hand, at ${t''=\frac{1}{2}}$, the dispersion near the $\mb K/\mb K'$ points is isotropic and quadratic, hence the dynamical critical exponent ${z=2}$.

\subsection{Gaplessness of general symmetric fermion model}\label{gaplessApp}

In this subsection, we will prove the statement in the main text that the general $\otwo$ + $\sfT$ + lattice reflection symmetric fermionic Hamiltonian ${H_\text{gen} = \sum_\y C_{|\y+\bm{\del}|}H_\y}$ (see~\eqref{dH0}) is always gapless at the $\mb{K}$ and $\mb{K}'$ points. 
In terms of the momentum space operators $c_\mathrm{W}(\k)$ and $c_{\mathrm{B}}(\k)$, the Hamiltonian $H_\text{gen}$ becomes
\begin{equation}
    \sum_{\k \in \bz}
\left(\t\Del_\k\, c^\dag_\mathrm{W}(\k) c_{\mathrm{B}}(\k) + 
\t\Del^*_\k \, c^\dag_{\mathrm{B}}(\k)c_\mathrm{W}(\k) \right),
\end{equation}
where ${\t\Del_\k = \sum_{\y} C_{|\y+\bm{\del}|} e^{i\k \cdot (\y+\bm{\del})}}$. Therefore, there are two bands with the single-particle dispersions ${+ |\t\Del_\k|}$ and ${- |\t\Del_\k|}$, respectively, and gapless modes will exist wherever ${\t\Del_\k = 0}$.

Since the Hamiltonian $H_\text{gen}$ has the entire crystalline symmetry of the honeycomb lattice, $\t\Del_\k$ can be rewritten as
\begin{equation}
    \t\Del_\k \!=\! \frac13\sum_{\y} C_{|\y+\bm{\del}|} (e^{i\k \cdot (\y+\bm{\del})}+e^{i\k \cdot U_{\mathrm{W}}(\y+\bm{\del})}+e^{i\k \cdot U_{\mathrm{W}}^2(\y+\bm{\del})}),
\end{equation}
where $U_{\mathrm{W}}$ is a ${2\pi/3}$ site centered rotation (see Eq.~\ref{UWandUBdef}). At the $\mb{K}$ and $\mb{K}'$ points, for all Bravais lattice vectors $\y$, 
\begin{align}
    \left(e^{i\k \cdot (\y+\bm{\del})}+e^{i\k \cdot U_{\mathrm{W}}(\y+\bm{\del})}+e^{i\k \cdot U_{\mathrm{W}}^2(\y+\bm{\del})}\right)\bigg\rvert_{\k = \mb{K},\mb{K}'} = 0.
\end{align}
Therefore, ${\t\Del_{\mb{K}} = \t\Del_{\mb{K}'} = 0}$, and the Hamiltonian $H_\text{gen}$ is always gapless as claimed. Furthermore, it is straightforward to check that, for small $\k$, ${|\t{\Del}_{\mb{K} + \k}|\propto |\k|}$ and ${|\t{\Del}_{\mb{K}' + \k}|\propto |\k|}$ where the multiplicative factors on $|\k|$ are $\k$ independent. Therefore, not only is $H_\text{gen}$ always gapless, but it always has Dirac cones at the $\mb{K}$ and $\mb{K}'$ points.

\def\bibsection{\section*{\refname}}
\bibliographystyle{apsrev4-1}
\bibliography{refs}

\end{document}